# Spin-photon Qubits for Scalable Quantum Network

Md Sakibul Islam[1], Kuldeep Singh[1], Yunhe Zhao[1], Nitesh Singh[1], Wayesh Qarony[1,2,3,*]

[1]*Department of Electrical and Computer Engineering, University of Central Florida, Orlando, FL 32816, USA*

[2]*Department of Physics, University of Central Florida, Orlando, FL 32816, USA*

[3]*CREOL, The College of Optics and Photonics, University of Central Florida, Orlando, FL 32816, USA*

*Corresponding Email: wayesh@ucf.edu

Solid-state quantum light sources offer a scalable pathway for interfacing stationary spin qubits with flying photonic qubits, forming the backbone of future quantum networks. Telecom-band spin-photonic qubits, operating in the 1260-1675 nm wavelength range, are particularly well-suited for long-distance quantum communication due to minimal loss in standard optical fibers. Achieving scalability, however, hinges on fulfilling several stringent criteria: coherent spin-state control, deterministic and indistinguishable single-photon emission, and integration with nanophotonic structures that enhance radiative properties, such as lifetime, coherence, and photon indistinguishability. This study explores the state-of-the-art spin-photonic qubits across solid-state platforms, including diamond color centers, silicon carbide defect centers, quantum dots, and two-dimensional materials. Special attention is given to silicon-based emitters, particularly G, T, C- and $C_i$-centers, which promise monolithic integration with complementary metal-oxide-semiconductor (CMOS) technology and telecom-band operation. We classify these systems based on spin-photon interface availability, CMOS process compatibility, and emitter scalability. We also discuss recent advances in cavity quantum electrodynamics (cQED), including Purcell enhancement and quality factor engineering in integrated photonic (circuits) environments. The work highlights emerging demonstrations of quantum networking over metropolitan scales and outlines the trajectory toward chip-scale quantum photonic integrated circuits (QPICs). It combines deterministic emitter creation, coherent spin manipulation, and quantum information processing. These developments pave the way for global quantum networks, enabling secure communication, distributed quantum computing, and quantum-enhanced sensing.

## 1. Introduction

Building a scalable quantum network requires the ability to reliably generate, manipulate, and distribute entangled quantum states between remote nodes. A foundational element in this architecture is the spin-photonic qubit, a hybrid quantum system where a stationary spin qubit serves as a local quantum memory and a flying photonic qubit transmits information across optical links.[1] The ability to interface spin and photonic degrees of freedom unlocks key functionalities, including quantum state transfer, entanglement distribution, and remote quantum gate operations.[2–6] Solid-state quantum emitters are well-suited for realizing spin-photonic qubits.[7,8] These systems offer atom-like energy level structures with optical transitions, enabling spin-state preparation and optical readout, while leveraging the scalability and integration potential of semiconductor materials.[9–13] Crucially, spin-photonic qubits must fulfill stringent requirements: (i) long spin coherence times, [13–16] (ii) deterministic generation of indistinguishable single photons, [17–19] (iii) efficient spin-photon interface coupling, [20–22] and (iv) compatibility with photonic integration and large-scale fabrication processes. [13,15,16,23]

**Figure 1** illustrates the operating principle of spin-photon qubits in the context of a scalable quantum network. A localized electronic spin in a solid-state host material is entangled with an emitted single photon via optical excitation and relaxation pathways.[24] This spin-photon entanglement forms the basis for quantum communication protocols, such as quantum teleportation and entanglement swapping,[24] and it is essential for establishing entangled links between quantum network nodes.[25,26] For long-distance quantum communication, the telecom wavelength band (1260-1675 nm) is highly advantageous due to the minimal attenuation in optical fibers, particularly near 1310 nm and 1550 nm, where losses are ~0.35 dB/km and ~0.14 dB/km, respectively.[27–30] Achieving single photon emission in this band requires quantum emitters with sub-bandgap transitions (~0.7-1.1 eV), often necessitating deep-level defect states or engineered semiconductor heterostructures.[31–34] To increase the indistinguishability and brightness of single photons, emitters are often integrated with optical cavities or photonic crystal structures.[19,35–37] The integrated nanophotonic structures enhance emission characteristics like Purcell factors with a narrow linewidth.[38–43] Furthermore, coherent control of the spin states via RF or microwave pulses enables manipulation of spin sublevels to perform quantum logic operations and memory functions.[44–47]

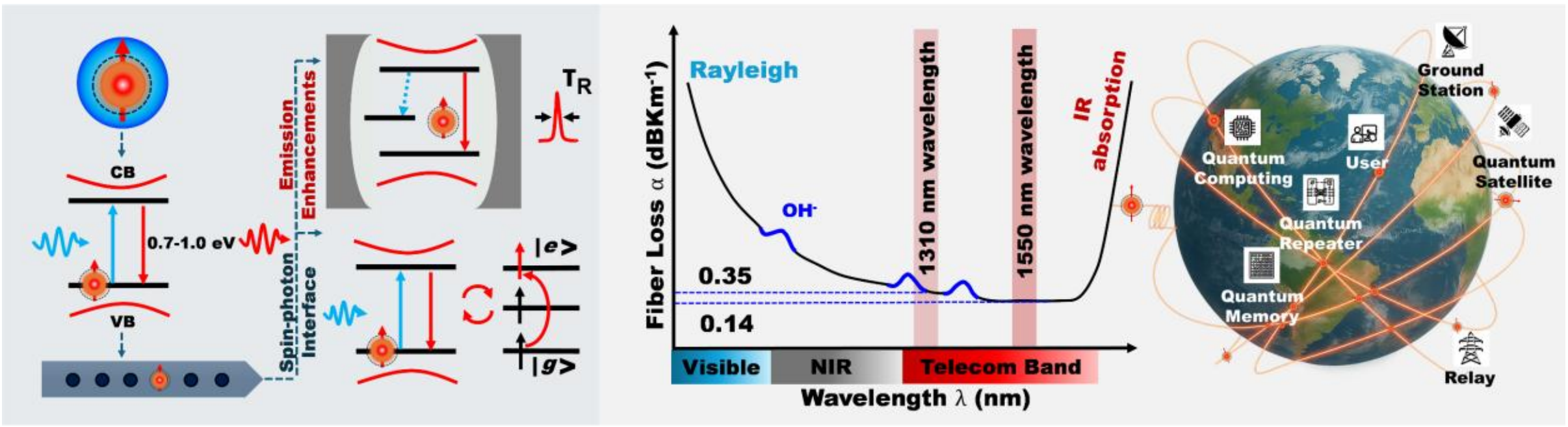

**Figure 1: Spin-photonic qubits for scalable quantum networks.** Spin-photonic qubits link a localized spin qubit (stationary quantum memory) to a flying photonic qubit (for long-distance transmission), enabling entanglement distribution across quantum networks. For telecom-band operation (1260-1675 nm), the emitter's optical transition energy must be in the ~0.7-1.0 eV range. Optical or electrical excitation drives an electronic transition from the ground to the excited state, while the spin degree of freedom remains addressable for initialization and readout through spin-dependent optical selection rules. Coupling to nanophotonic cavities enhances emission via high Purcell factor, reducing radiative lifetimes, narrowing linewidths, and increasing photon collection efficiencies. Coherent RF or microwave control allows manipulation of spin sublevels and deterministic emission of entangled photons while preserving long-lived spin memories. Telecom wavelengths at 1310 nm and 1550 nm offer ultra-low fiber attenuation (~0.35 and ~0.14 dB/km, respectively), outperforming visible and mid-IR ranges that suffer from Rayleigh scattering or infrared absorption. Leveraging these low-loss telecom windows, spin-photon interfaces in silicon can be integrated into fiber and satellite links to realize large-scale quantum connectivity, enabling distributed quantum computing, quantum repeaters, global quantum key distribution, and secure quantum communications.

The diverse landscape of solid-state quantum emitters is illustrated in **Figure 2**. It maps a range of platforms across the visible (green, 520-780 nm), near-infrared (NIR, blue, ~790-1100 nm), and telecom (magenta, 1260-1675 nm) spectral bands. The different spectral regimes are color-coded by emission wavelength and are radially segmented by scalability. The figure highlights the trade-offs between photon emission properties, device integration maturity, and suitability for quantum network deployment. Emitters in the visible range, such as quantum dots (QDs), group IV color centers in diamond-based systems, typically offer high photon purity and brightness but suffer from high fiber attenuation in the visible and require quantum frequency conversion for a long-distance quantum network.[48,49] Although NIR-emitting platforms such as silicon-vacancy centers ($V_{Si}$), vanadium-based defects ($V^{4+}$), and indium arsenide (InAs) based QDs have better spin coherence property and low loss relative to the visible range, they are still suboptimal for long-haul fiber-based communication.[48] In contrast, telecom-band emitters, such as those based in silicon (e.g., C, $C_i$, T, and G centers), and rare-earth ions like $Er^{3+}$, are highly attractive due to their spectral alignment with fiber transmission maxima, long optical and spin coherence time, and potential compatibility with CMOS photonic integration.[50–52] Notably, silicon-based spin-photon qubits, including G center, T center, $C_i$ center, and C center have emerged as particularly promising candidates for telecom-band operation, with zero-phonon lines (ZPLs) at 1280 nm, 1326 nm, 1452 nm, and 1571 nm, respectively.[53–57] The ZPLs in the low-loss spectral window of fibers, along with the industrial scalability of silicon, make these emitters a strong contender for realizing CMOS-compatible quantum networks.

Silicon's appeal extends beyond its spectral compatibility. It has well-established fabrication infrastructure, low optical losses, and compatibility with cryogenic operation,[58] which make it an ideal platform for scalable quantum technologies.[50,59,60] Silicon-based emitters have demonstrated spin-selective optical transitions in the telecom band, and long spin coherence time.[32,55] Furthermore, the potential for deterministic creation of emitters via ion implantation or laser annealing opens the door to integrating these

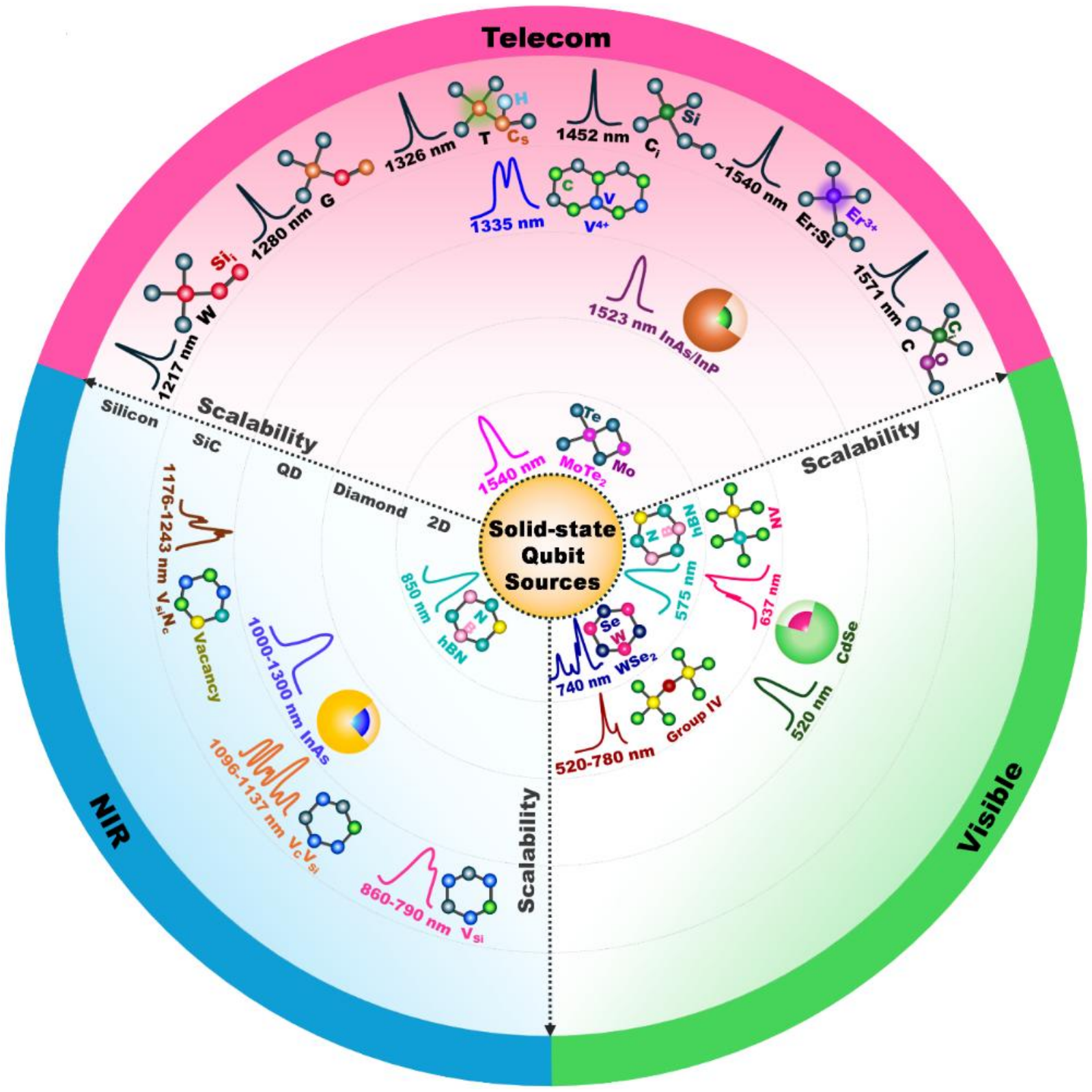

**Figure 2: Single-photon emission across solid-state quantum emitters.** The solid-state single-photon sources span the visible (400-780 nm), near-infrared (NIR, ~790-1100 nm), and telecom (1260-1675 nm) spectral ranges. This radial diagram categorizes emitters by wavelength (outer color ring: green = visible, blue = NIR, magenta = telecom) and scalability (radial direction, center = emerging platforms, outer edge = highly scalable). Scalability is evaluated based on telecom band emission, availability of spin–photon interfaces, and CMOS-compatible fabrication maturity. Silicon (Si) stands out as a unique host with multiple defect centers, including W, G, T, Ci, and C, that emit in the telecom band. The C center (interstitial carbon + substitutional oxygen) emits at 1571 nm, close to the lowest fiber loss window. The T center (two substitutional carbons + hydrogen) offers spin–photon entanglement and potential for electrically driven operation. The Ci center (~1452 nm) is a minimal-carbon defect with narrow optical lines and potential spin addressability. Together, these centers position silicon as a leading platform for scalable quantum networks, leveraging mature foundry infrastructure and full CMOS compatibility. Other telecom emitters include $V^{4+}$ centers in SiC, InAs/InP quantum dots, and select 2D materials. Most SiC color centers (e.g., divacancies) emit in the NIR, while diamond NV and SiV/SnV centers are highly advanced for quantum memory and metropolitan networking, but lack intrinsic telecom emission. Representative 2D emitters such as hBN (~850 nm) and $WSe_2$ (~740 nm) can benefit from photonic integration (e.g., circular Bragg gratings) to enhance brightness and collection efficiency.

centers with silicon photonic components such as waveguides, micro-ring resonators, and nanocavities. This monolithic integration not only facilitates efficient photon routing and filtering but also enables

Purcell-enhanced emission. In addition, it improves photon indistinguishability and provides a scalable packaging for quantum networks.[61,62] Therefore, silicon-based spin-photonic qubits represent a key step toward building quantum photonic integrated circuits (QPICs) capable of large-scale quantum state generation, entanglement distribution, and on-chip quantum processing.[23,59,63–65]

In this study, we present a comprehensive survey of solid-state spin-photonic qubits, with a particular focus on their potential to enable scalable quantum networks. We assess leading emitter platforms, including diamond, QD, and silicon carbide (SiC), using key performance metrics such as emission wavelength, photon indistinguishability, spin coherence properties, and control mechanisms. In terms of fabrication, the metrics are scalability, deterministic placement, and compatibility with photonic circuitry and CMOS process integration. Special emphasis is placed on recent advances in telecom-band silicon-based color centers, which offer long spin coherence times and seamless integration with foundry-compatible photonic platforms. We also highlight demonstrations of quantum networking protocols, such as Bell-state generation, entanglement swapping, quantum key distribution, and metropolitan-scale entanglement distribution. Furthermore, we present the progress in on-chip spin-photon interfaces and nanophotonic integration strategies that include Purcell enhancement, deterministic emitter-cavity coupling, and cavity quantum electrodynamics (cQED), all of which are critical for realizing QPICs. As depicted in recent prototype systems, these integrated platforms hold promises to converge the functionalities of quantum memory, photon routing, modulation, and detection onto a single chip. Together, these advances lay the groundwork for practical implementations of distributed quantum information processing, secure communication, and large-scale quantum network infrastructure.

## 2. Key Concepts in Spin-Photonic Qubits

The optical performance of spin-photonic qubits can be significantly improved by integrating them into nanophotonic structures that modify the emitter's photonic environment (**Figure 3**). The Purcell effect, which enhances the spontaneous emission rate by engineering the local density of optical states, plays a central role in improving photon extraction efficiency and indistinguishability.[66,67] The Purcell factor, FP $\propto Q/V$, scales with the cavity quality factor $Q$ and inversely with the mode volume $V$. Photonic crystal cavities, ring resonators, and nanobeam waveguides are frequently employed to achieve strong emitter-cavity coupling.[68,69]

Photonic crystal cavities, in particular, are engineered by introducing a periodic dielectric structure with a photonic bandgap that prohibits light propagation in specific frequency ranges.[70–73] Introducing a defect, such as removing or displacing photonic crystal holes, creates a localized mode within the bandgap, confining light to an ultra-small volume with minimal loss.[7475,76] When spin-photonic qubits are placed at the maximum field of such structures, they can emit photons into well-defined spatial and spectral modes with greatly enhanced radiative rates. This deterministic coupling improves brightness, reduces emission into lossy channels, and enhances collection efficiency for quantum photonic circuit integration.

Realizing scalable spin-photonic qubits in solid-state platforms requires the ability to generate, characterize, and coherently manipulate single photons entangled with localized spin states.[77–80] This section provides an overview of essential experimental schemes and physical principles that characterize the performance and evaluation of spin-photonic qubits in integrated systems.

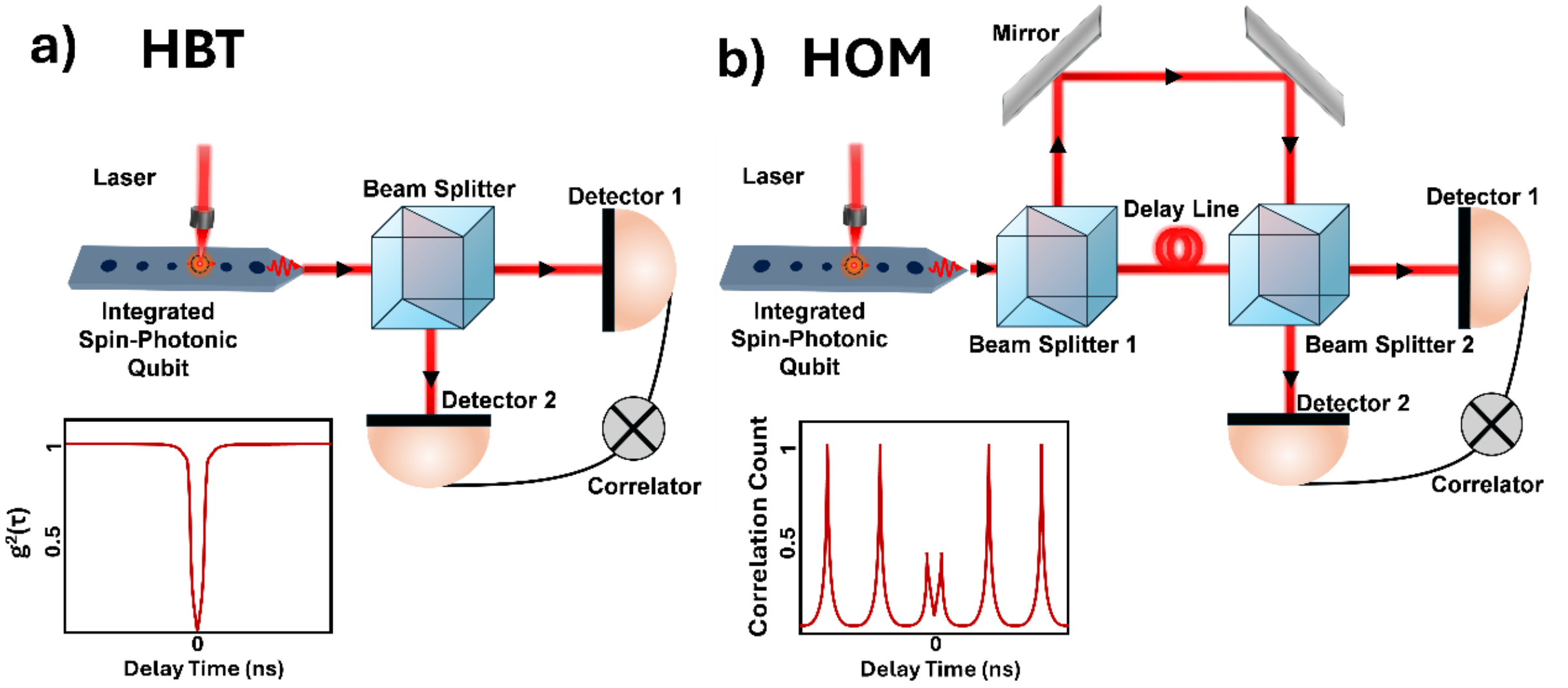

**Figure 3: Optical characterization of integrated spin-photonic qubit:** (a) The Hanbury Brown-Twiss (HBT) experiment measures the second-order correlation function, $g^2(0)$, by splitting the emitted photons with a beam splitter (50:50) and detecting them with two detectors to confirm single photon emission through antibunching. (b) The Hong-Ou-Mandel (HOM) experiment extends this HBT scheme with a delay line and a second beam splitter, enabling two-photon interference to validate indistinguishability of photons.

**Photon Statistics and Quantum Light Characterization**. Deterministically generated single photons from quantum emitters should exhibit high brightness, spectral purity, photon indistinguishability, and long coherence times.[81] The two essential signatures of quantum light are antibunching and indistinguishability, which are typically evaluated through Hanbury-Brown and Twiss (HBT) and Hong-Ou-Mandel (HOM) experimental schemes, respectively (**Figure 3**).[82–84] In an HBT experiment, the second-order correlation function, $g^2(0)$, is measured by directing photons from the emitter through a 50:50 beam splitter and detecting correlated coincidences at the two output ports (**Figure 3(a)**). The value of $g^2(0)$, the probability of detecting two photons at zero-time delay, serves as a quantitative measure of photon antibunching. For an ideal single-photon source $g^2(0) = 0$, indicating the complete suppression of multi-photon emission. Experimentally, the value of $g^2(0) < 0.5$ is considered as strong evidence of single-photon behavior.[85] The HBT histogram also provides insights into the emitter's lifetime and emission dynamics. Therefore, the low $g^2(0)$ values and narrow temporal features indicate high-purity quantum emitters with fast recombination rates.

The indistinguishability of single photons is characterized using the HOM interference experiment. In this experiment, two photons, emitted either from the same source at different times or from two identical sources, are temporally overlapped using a beam splitter (**Figure 3(b)**). The perfectly indistinguishable photons interfere destructively and exit from the same output port, producing a dip in coincidence counts, known as the HOM dip.[86] The visibility of this dip reflects how identical the photons are across all degrees of freedom, spectral, temporal, spatial, and polarization.[87–91] The high-visibility interference requires narrow ZPLs with minimal spectral broadening, which is critical for protocols such as entanglement swapping and photonic quantum teleportation in quantum networks.

**Spin Initialization and Coherent Control.** The ability to coherently manipulate and read out spin states is central to realizing quantum memories, performing gate operations, and generating entanglement between matter and light.[92–94] For practical quantum information processing, spin qubits must exhibit long coherence times, quantified by the dephasing time ($T_2^*$) and spin-echo coherence time ($T_2$), to allow for initialization, control, and sequential gate operations. Achieving long coherence time requires host materials with low magnetic noise, ideally with isotopically purified environments that reduce hyperfine interactions and minimize paramagnetic impurities.[95–97] Platforms such as diamond, silicon, and SiC have successfully employed isotopes like $^{12}C$ or $^{28}Si$ to enhance spin coherence and suppress decoherence pathways.[98,99] However, spin-lattice relaxation (governed by $T_1$) still imposes temperature-dependent limitations, often necessitating cryogenic operation to suppress phonon-mediated processes.[100]

The coherent spin control is typically achieved via optically detected magnetic resonance (ODMR), which enables initialization, microwave manipulation, and optical readout of spin states.[101] A microwave field resonant with the spin transition drives coherent population transfer, and changes in photoluminescence (PL) intensity as a function of microwave frequency yield the ODMR spectrum. The spectrum reveals spin transition frequencies, zero-field splitting, and hyperfine coupling constants, which are the key parameters of the spin Hamiltonian.[102] The ODMR has been essential in demonstrating long-lived coherence and quantum control in NV and SiV centers in diamond, hBN in 2D, as well as the G and T center in silicon, which hosts both an electron spin and a hyperfine-coupled nuclear spin suitable for spin registers and memory storage.[55,102–109] The combination of spin initialization, coherent driving, and optical readout lays the foundation for quantum logic operations and entanglement protocols in spin-photonic systems.

**Spin-Photon Entanglement Protocols.** Demonstrating entanglement between a localized spin qubit and a flying photonic qubit is a foundational step toward building quantum networks.[80,110] In solid-state systems, this requires control over the optical selection rules that govern spin-conserving transitions, as well as precise preparation and readout of the spin state. The defect centers with spin-triplet ground states and optically addressable excited states, such as NV and SiV centers in diamond, hBN in 2D, or the G/T centers in silicon, are particularly suitable for these protocols due to their spin-selective radiative pathways.[50,55,107,109,111–113] In a typical spin-photon entanglement experiment, the spin is initialized into a coherent superposition of ground states using a resonant microwave or RF pulse. A resonant optical pulse then excites the system to a spin-conserving excited state. The subsequent emission results in a photon whose properties, such as polarization, frequency, or emission time, carry information about the spin qubit and serve as the quantum link to a distant node.[49,114]

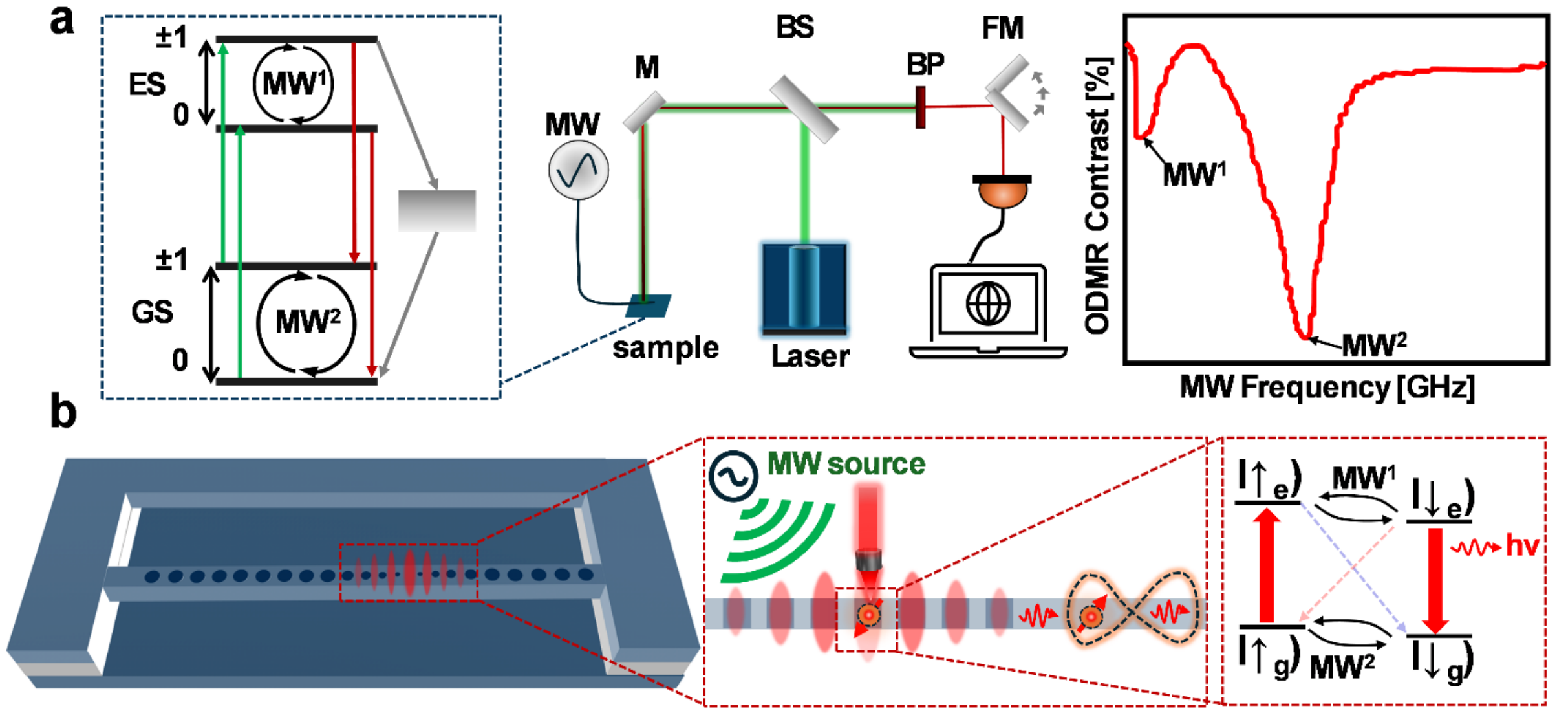

**Figure 4: Spin control and entanglement protocols:** Demonstrating spin control and entanglement between a localized spin qubit and a photonic qubit. (a) The spin qubit is initialized into a coherent superposition of ground or excited states using microwave pulses (MW). Subsequent spin-photon transition driven by microwave excitation is collected through an optical experimental setup consists of mirror (M), beam splitter (BS), bandpass filer (B) and detectors known as Optically Detected Magnetic Resonance (ODMR). (b) A schematic of spin-photon entanglement protocols for monolithically integrated nanophotonic structures is shown. Inset shows optical excitation of a spin-photonic qubit and its manipulation using microwave source results in spin-selective transition. The emitted photon as flying qubit carries information for an entangled state.

The verification of spin-photon entanglement requires joint measurements on both the spin and photonic degrees of freedom. Typically, spin-selective photoluminescence contrast yields the ODMR spectra (**Figure 4**).[114] The high-visibility correlations confirm the entanglement and enable the implementation of quantum protocols such as entanglement swapping, quantum teleportation, and remote spin-spin entanglement.[115,116] Notably, such protocols have been demonstrated using NV centers over kilometer-scale fiber links (**Figure 6a**),[117] SiV centers in integrated nanophotonics across metropolitan distances (**Figure 6b**),[49] and III-V QDs in free-space and fiber networks (**Figure 6c**).[78] These developments provide essential building blocks for quantum repeater architectures and multi-node quantum networks based on integrated spin-photonic interfaces.

## 3. Integrable Solid-State Qubit Platforms for Quantum Networks

Solid-state spin-photonic qubits have emerged as a central building block for scalable quantum networks by enabling the generation of on-demand, indistinguishable, and entangled single photons from compact, chip-integrated systems.[118–120] These platforms range from defect centers in wide-bandgap crystals to semiconductor QDs and atomically thin 2D materials. The combination of discrete energy levels, spin-selective optical transitions, and nanofabrication compatibility makes them well-suited for integration with photonic circuits and control electronics.[121–123] In addition, their feasibility for optical or electrical excitation and compatibility with mature fabrication processes support scalable quantum information processing, both in free-space and on-chip architectures.[50,124] This section presents a comprehensive study of solid-state spin-photonic qubit platforms, focusing on their emission properties and nanophotonic integration. The discussion is organized around five principal material systems: (i) diamond, a wide-bandgap host for optically stable and long-lived defect spins; (ii) III–V QDs, offering deterministic photon generation with scalable epitaxial fabrication; (iii) silicon carbide (SiC), which combines telecom-band emission with CMOS process compatibility; (iv) two-dimensional (2D) materials, where tunable excitonic and defect-based emission emerges from atomically confined layers and (v) silicon, a technologically dominant platform now showing promise as a host for telecom-wavelength color centers. Each of the systems is evaluated based on a consistent set of criteria, including emission wavelength (particularly proximity to telecom bands spanning), ZPL fraction, spin coherence times and optical lifetimes, photon statistics such as second-order correlation, $g^2(0)$, and indistinguishability. We also evaluated the system based upon their demonstration of integration with photonic cavities, waveguides, and modulators. Moreover, we examined each platform's potential for deterministic emitter placement and CMOS-compatible fabrication, along with key demonstrations of spin-photon entanglement, Bell-state generation, and quantum teleportation. This framework provides a unified perspective on current progress and challenges in developing spin-photonic qubits for scalable quantum technologies.

### 3.1 Diamond Defect Centers

Diamond has long served as a foundational material platform for solid-state quantum technologies, owing to its exceptional optical transparency, ultrawide electronic bandgap (5.5 eV), and low nuclear spin density in isotopically purified $^{12}C$.[125] These properties make it an ideal host for optically active defect centers with well-isolated electronic states and long-spin coherence times.[126–129] Over the past two decades, diamond has enabled key milestones in quantum information science, including spin-photon entanglement, quantum memory protocols, and the first metropolitan-scale quantum network demonstrations.[80,117,130,131] Among the various defect centers, the nitrogen-vacancy (NV) [117,132,133] and group-IV vacancy centers,[134] particularly silicon vacancy (SiV) [115] and tin vacancy (SnV),[135,136] have emerged as leading candidates for spin-photonic qubits.

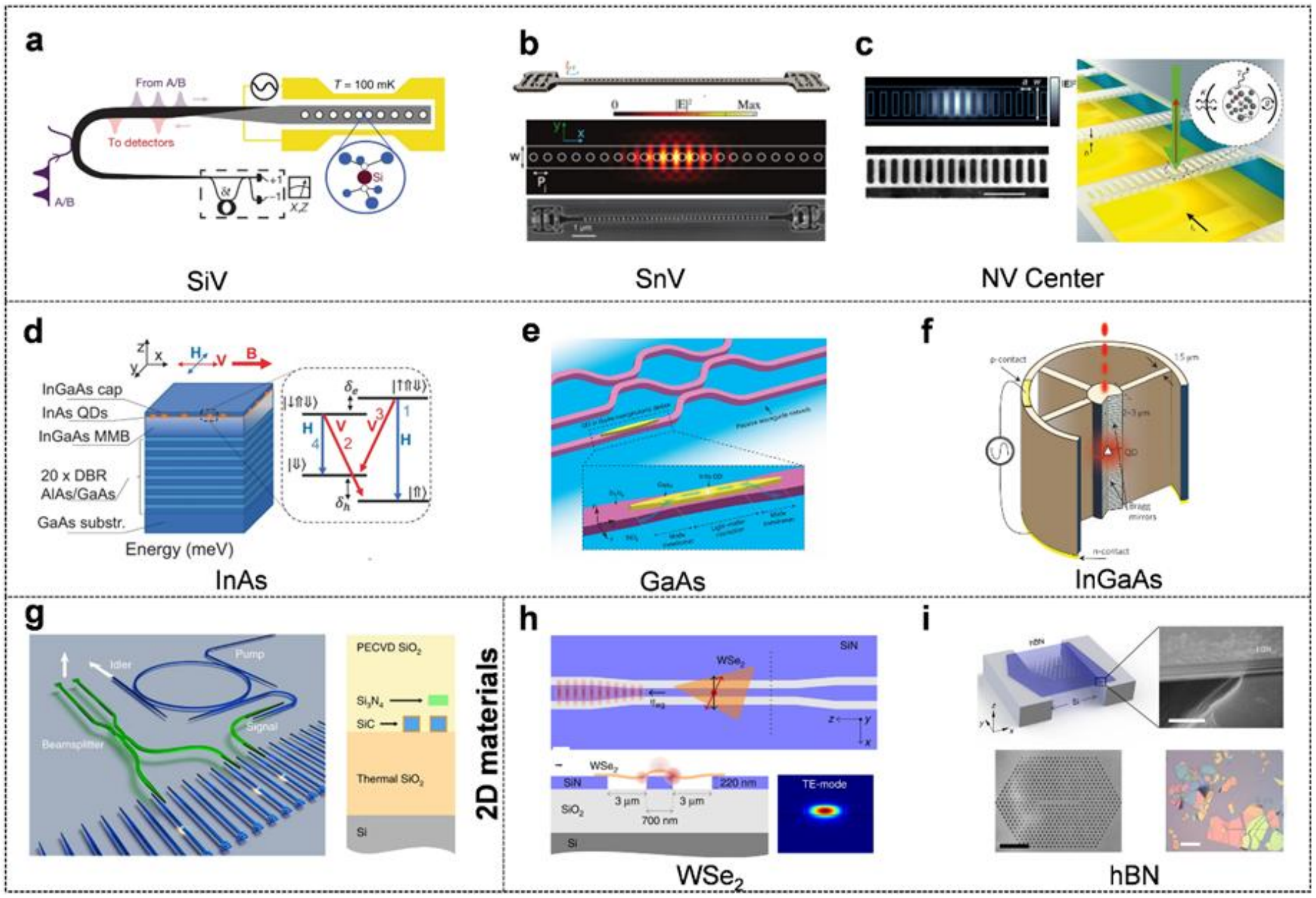

**Figure 5: On-chip spin-photon interfaces across solid-state quantum emitters.** Single-photon emission from solid-state quantum emitters, including diamond color centers, quantum dots, silicon carbide (SiC) defects, and two-dimensional (2D) materials, has been demonstrated in integrated photonic platforms. On-chip spin-photon interfaces enable efficient coupling between long-lived spin qubits (quantum memories) and photonic flying qubits, enhancing entanglement distribution for scalable quantum networks. (a-c) Optically addressable nanophotonic cavities or waveguides integrating diamond silicon-vacancy (SiV), tin-vacancy (SnV), and nitrogen-vacancy (NV) centers for coherent spin control and enhanced photon collection (a) Adapted from ref.[131] ©Springer Nature, (b) Adapted from ref. [160]; licensed under a Creative Commons Attribution (CC BY), and (c) Adapted from ref.[159]; licensed under a Creative Commons Attribution (CC BY)) (d) InAs quantum dots enabling spin-selective optical transitions with emission in the telecom band (Adapted from ref.[168] ; licensed under a Creative Commons Attribution (CC BY)). (e) GaAs quantum dots embedded in micro-ring resonators and bus waveguides for efficient fiber-compatible photon extraction (Adapted from ref.[169]); licensed under a Creative Commons Attribution (CC BY)). (f) Electrically injected InGaAs quantum dots in planar microcavities generate near-unity indistinguishable single photons (Adapted from ref. [165]) ©Springer Nature). (g) Monolithic 4H-SiC platforms with frequency-converted silicon-vacancy emission, compatible with telecom-band quantum networking (Adapted from ref.[312]) ©Springer Nature). (h-i) Integrated photonic crystal cavities coupled to $WSe_2$ and hBN emitters, demonstrating cavity quantum electrodynamics (cQED) effects to boost brightness, directionality, and spectral purity ((h) Adapted from ref. [251]; licensed under a Creative Commons Attribution (CC BY) and (i) Adapted from ref.[76] ); licensed under a Creative Commons Attribution (CC BY)).

The negatively charged NV center consists of a substitutional nitrogen atom adjacent to a vacant carbon site, forming a defect with $C_{3v}$ symmetry. Its electronic structure features a spin-triplet ground state and an optically addressable spin-triplet excited state, separated by a ZPL at 637 nm.[125,137] The NV center supports full spin qubit control: optical pumping at 532 nm initializes the spin state and coherent manipulation with microwave sources enables spin-selective optical transitions, while spin readout is achieved through photoluminescence contrast.[138,139] Additionally, the NV's electronic spin can be coherently coupled to nearby $^{13}C$ nuclear spins or the host $^{14}N$ nucleus, forming multi-qubit registers with coherence times exceeding one second under dynamical decoupling.[140–145] Despite advantages, NV centers suffer from optical limitations that hinder their scalability in photonic quantum networks.[146] Most notably, only ~3-5% of emission occurs in the ZPL, with the remainder distributed in a broad phonon sideband, reducing photon indistinguishability.[144,147] Moreover, the lack of inversion symmetry makes the NV center highly sensitive to electric field fluctuations via the Stark effect, resulting in spectral diffusion and inhomogeneous broadening.[148–150] These effects complicate the generation of indistinguishable photons from separate NV centers, an essential requirement for photon-mediated entanglement protocols.

To overcome the limitations of the NV center, recent research has focused on group-IV vacancy centers, which exhibit inversion symmetry and superior optical coherence. The SiV center consists of a silicon atom positioned symmetrically between two adjacent carbon vacancies along the ⟨111⟩ axis, forming a split-vacancy configuration with $D_{3d}$ symmetry.[148] This inversion symmetry eliminates the permanent electric dipole moment, making the optical transitions of the SiV center inherently robust to local electric field noise.[151,152] As a result, SiV centers show narrow homogeneous linewidths and spectral stability even in nanofabricated environments.[153] The SiV center emits a ZPL at 737 nm wavelength with a Debye-Waller (DW) factor of ~70%, meaning that most of its photons end up in the ZPL.[154,155] This property contrasts sharply with the NV center and enables high-fidelity spin-photon interfaces. The spin coherence time of SiV centers, however, is typically limited to microseconds at 4 K due to phonon-mediated orbital relaxation, but improves significantly at sub-kelvin temperatures (<100 mK) where spin $T_2$ times up to tens of milliseconds have been reported.[105,156] The SnV center offers a similar structure but with a larger ground-state splitting (~850 GHz), suppressing phonon-induced decoherence and improving thermal stability.[136] It makes SnV a promising candidate for operation in the 1-4 K regime without dilution refrigeration.

**Nanophotonic Integration and Emission Enhancement.** Integrating diamond defect centers into nanophotonic cavities is essential for enhancing light-matter interactions and enabling scalable quantum photonic circuits.[157,158] Among various architectures, one-dimensional nanobeam cavities have shown particular success in improving emission from centers (**Figure 5a-c**).[131,159,160] These structures consist of a suspended diamond waveguide patterned with a periodic array of air holes, forming a photonic crystal with a localized defect that traps optical modes in a small optical mode volume V with high Q quality factors.[158] In a particular case, SiV centers coupling to such cavities has led to a 42-fold enhancement of ZPL intensity, 10-fold reduction in excited state lifetime, and a Purcell factor exceeding 10.[158] These advances improve brightness and indistinguishability and allow for deterministic photon routing into on-chip waveguides or fiber-coupled systems.[49] Beyond nanobeams, two-dimensional photonic crystal slabs and micro-ring resonators have also been explored, supporting slow-light modes that increase light-matter interaction time and broaden the operational bandwidth for inhomogeneous emitter ensembles. Efforts to integrate NV centers with photonic cavities have also been successful in partially compensating for their low intrinsic ZPL fraction.[157] The cavity enhancements of up to 7-fold have been demonstrated, improving photon extraction efficiency and enabling interference-based quantum protocols at cryogenic temperatures.[157]

**Quantum Network Demonstrations in Diamond Defect Centers.** A milestone experiment conducted at TU Delft that connects two independent NV-based quantum nodes using 1.3 km of deployed fiber on the campus. The spin-spin entanglement between the nodes was achieved through heralded single-photon detection, and photons were converted from 637 nm to the telecom band using quantum frequency conversion to enable long-distance transmission (**Figure 6a**).[161] The setup included active path-length stabilization, entanglement heralding, and storage in long-lived nuclear spin registers. Recently, SiV centers integrated into nanophotonic cavities were used in a 35 km metropolitan network in Boston (**Figure 6b**).[49] Another key advance was achieved using $SnV^{-}$ centers in diamond, which exhibit strong optical transitions and enhanced thermal stability due to their large ground-state splitting (~850 GHz). In a recent experiment (**Figure 6d**), individual SnV emitters in separate cryostats were linked by a 70 m polarization-maintaining fiber, and local electrostatic tuning was used to bring their optical transitions (619 nm) into resonance.[162]

This two-photon Hong–Ou–Mandel interference with visibility approaching 80 % demonstrated high photon indistinguishability and remote quantum interference between solid-state emitters, representing an essential step toward multi-node diamond-based quantum networks. Although these demonstrations operated at visible wavelengths, quantum frequency conversion to the telecom band remains essential for truly long-distance, low-loss quantum networking. Collectively, these systems exploit fast optical transitions for efficient spin–photon entanglement and leverage nearby nuclear-spin registers (e.g., $^{29}Si$) as robust quantum memories, enabling high-fidelity entanglement generation, storage, and distribution across emerging quantum networks.

Diamond continues to be a leading platform for spin-photonic qubits, offering unrivaled spin coherence and demonstrated utility in quantum networking. However, challenges remain in scaling to large networks. The NV, SiV, and SnV centers require low-temperature operation to suppress phonon-mediated dephasing and spectral stabilization techniques to improve indistinguishability. Further development of deterministic fabrication, hybrid integration with silicon photonics, and frequency conversion technologies will be critical for leveraging diamond color centers in future chip-scale quantum architectures.

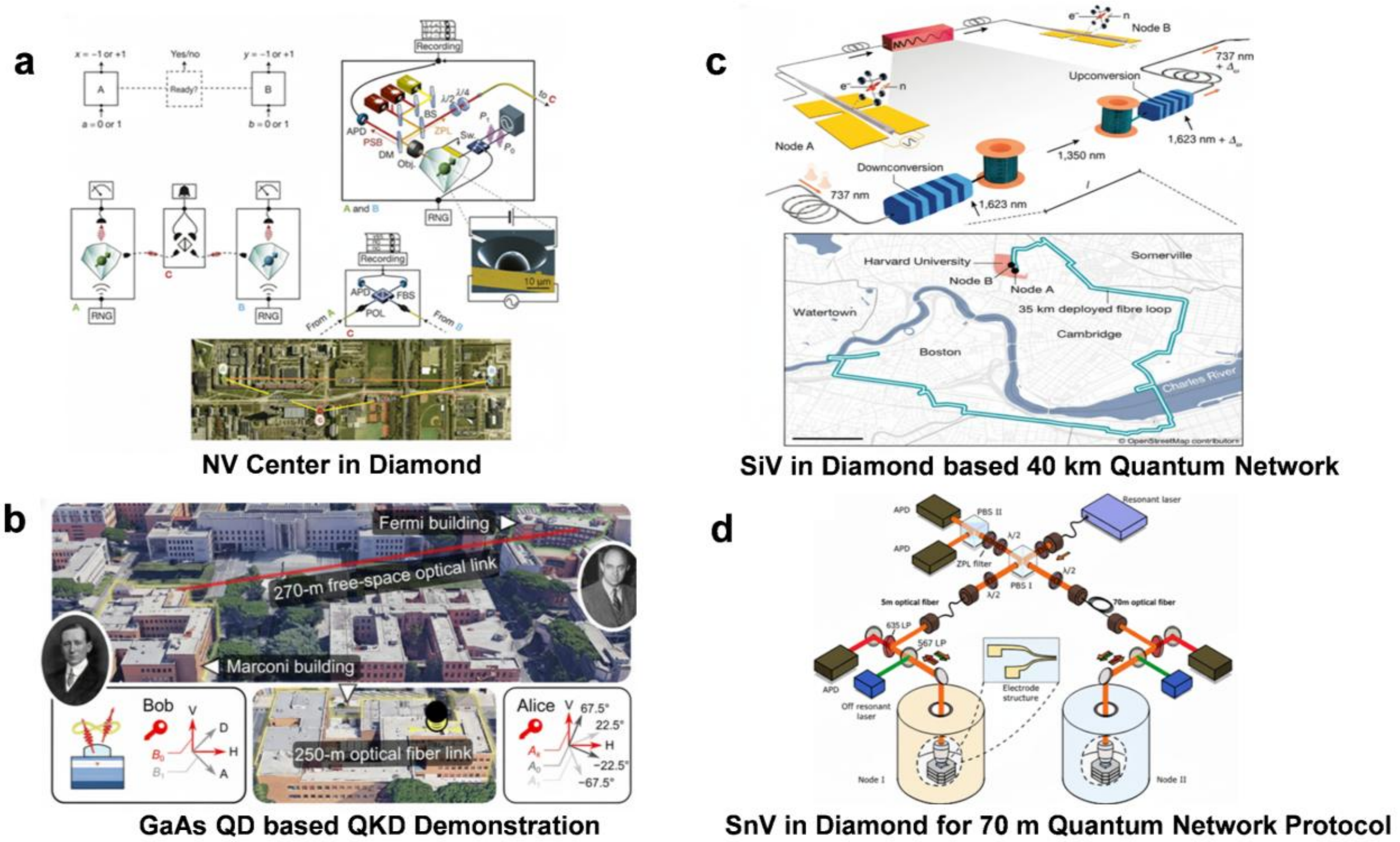

**Figure 6: Quantum network demonstrations in metropolitan areas.** (a) NV center in diamond: CHSH–Bell inequality test using electron spins of nitrogen-vacancy (NV) centers in diamond, separated by a 1.3 km deployed fiber link across the TU Delft University campus, Netherlands. Adapted from ref. [117] (©Springer Nature) (b) GaAs QDs: Entangled photon pairs from GaAs-based QDs enable quantum key distribution (QKD) between nodes at Sapienza University, Rome, via hybrid free-space (270 m) and fiber (250 m) optical links. Secret keys are generated by matched basis measurements at the sender and receiver. Adapted from ref. [78] (©Springer Nature) (c) SiV center in diamond: Two-node quantum network spanning 35 km in the Boston metropolitan area. Silicon-vacancy (SiV) centers integrated with nanophotonic devices enhance spin–photon entanglement, maintained at 1350 nm over the deployed fiber loop. Adapted from ref. [49] (©Springer Nature) (d) Two-photon interference between remote tin-vacancy ($SnV^-$) centers in diamond. Individual SnV emitters housed in separate cryostats were connected by a 70 m polarization-maintaining fiber. Local electric-field tuning brought their optical transitions (619 nm) into resonance, enabling Hong–Ou–Mandel interference with ≈ 80 % visibility. Adapted from ref. [162] (licensed under a Creative Commons Attribution (CC BY)). While the photons were not frequency-converted, such conversion to the telecom band will be essential for long-distance fiber-based quantum networking.

## 3.2 Quantum Dots (QDs)

The QDs are nanoscale semiconductor structures that confine electrons and holes in all three spatial dimensions, mimicking the discrete energy levels of atoms.[163] This artificial quantization enables QDs to function as highly tunable, on-demand single-photon sources with deterministic emission properties.

Among solid-state quantum emitters, III-V QDs, particularly InAs/GaAs systems grown via molecular beam epitaxy (MBE), are the most mature platforms for photonic quantum technologies (**Figure 5d-f**). Their ability to emit highly indistinguishable photons with near-unity efficiency and fast radiative lifetimes makes them a key building block for scalable quantum networks, especially in fiber-based and free-space optical architectures.[164–166]

The QDs are typically formed by self-assembled growth techniques, such as Stranski-Krastanov epitaxy, where lattice mismatch between layers induces strain-driven island formation.[167] The InAs embedded in a GaAs matrix is widely used due to its direct bandgap and emission wavelengths in the NIR range (~900-1000 nm). However, it can be extended into the telecom bands (1300-1550 nm) via advanced growth and strain engineering techniques (**Figure 5d–f**).[168–170] The emission wavelengths can be tuned further by adjusting dot size, composition, and the local electromagnetic environment. The QDs exhibit fast radiative lifetimes (typically 0.5-1 ns), allowing for GHz-rate single-photon generation.[166,171] Their emission can be triggered deterministically using resonant or quasi-resonant pulsed laser excitation, which reduces multiphoton contributions and spectral jitter.[172] Importantly, QDs can emit photons with high indistinguishability, primarily when operated under resonant excitation and embedded in photonic cavities that suppress phonon sidebands.

The QDs excel in generating high-purity single photons, routinely demonstrating $g^2(0) < 0.01$ in HBT measurements. Combined with resonant excitation and Purcell enhancement, these can produce nearly Fourier-transform-limited photons with indistinguishability exceeding 90%.[173–175] These characteristics have even enabled quantum interference in HOM experiments between remote dots tuned into resonance using electric or strain fields.[48]

The spin-photon interfaces can be realized by embedding single electron or hole spins in QDs.[176] The spin states can be initialized and manipulated via optical or microwave fields and entangled with emitted photons using polarization or time-bin degree of freedom. Charged QDs (i.e., QDs with a resident electron or hole) support spin-selective optical transitions in a $\Lambda$-type configuration, enabling coherent spin-photon entanglement protocols.[177–179] The optically addressable spin states serve as quantum memories and intermediate nodes for quantum repeaters.[176] However, maintaining long spin coherence in QDs remains challenging due to interactions with the nuclear spin bath of the host material (e.g., $^{69}$Ga, $^{75}$As), which induces hyperfine-mediated decoherence.[180–183] These effects can be mitigated by spin echo and dynamical decoupling techniques, extending coherence times to several microseconds. Still, values are shorter than those achievable in nuclear-spin-free hosts like diamond or SiC.[184]

**Integration with Photonic Nanostructures.** The III-V group QDs are compatible with monolithic integration into nanophotonic circuits, enabling efficient photon extraction and emission control. They have been embedded in a wide variety of structures, including micropillar cavities, photonic crystal cavities, and nanowire waveguides.[172,185–188] For instance, Purcell enhancement exceeding 10 has been achieved in QDs embedded in high-Q micropillars, resulting in faster emission rates, improved directionality, and indistinguishability.[189] One key advantage of QDs is their deterministic placement using site-controlled growth or in situ lithography techniques, allowing pre-determined integration into photonic structures.[163] Recent advances in hybrid integration have also enabled coupling QDs to silicon nitride and lithium niobate photonic circuits, expanding their applicability to CMOS-compatible platforms.[23,190,191]

**Telecom Band QDs and Metropolitan-Scale Networks.** The emission in the telecom bands is critical for quantum communication over long distances. However, most QD systems emit at the NIR spectral regime.[192] With some exceptions, recent studies show that QD devices based on InP are capable of electrically injected single photon emission in the telecom band.[193] To achieve emission at the telecom band, InAs/InP QDs grown on distributed Bragg reflector (DBR) based cavities, demonstrating emission near 1300 nm (O-band) and 1550 nm (C-band).[193–195] The quality of a single phone reported to have $g^2(0) < 0.02$ and entangled photon pairs was generated through the biexciton-exciton cascade.[186,196,197] Alternatively, frequency down-conversion of photons from near-infrared-emitting QDs to telecom wavelengths has been demonstrated using nonlinear crystals and periodically poled lithium niobate waveguides, preserving entanglement and indistinguishability.[193–195] These advances have led to fully functional QD-based quantum key distribution (QKD) systems and quantum repeater nodes operating over deployed optical fiber networks. Notably, entangled photon pairs generated from InAs QDs were used to demonstrate a free-space and fiber-based QKD system in Rome, validating their use in metropolitan-scale networks (**Figure 6c**).[198]

The QDs remain one of the most scalable and deterministic sources of single and entangled photons, with a clear roadmap toward integration in quantum network architectures. However, key challenges include improving spin coherence, minimizing charge noise and spectral diffusion, and expanding emission into the whole telecom window. The continued progress in material engineering, site-controlled growth, and hybrid integration with silicon photonics will position QDs as leading candidates for local and long-distance quantum communication.

### 3.3 Silicon Carbide (SiC)

SiC is a wide-bandgap semiconductor widely used in power electronics and high-temperature applications. Its mature wafer-scale fabrication infrastructure, compatibility with CMOS processing, and broad polytypic diversity make it an attractive host for scalable spin-photonic qubits.[199–201] Over the past decade, SiC has emerged as a versatile platform for solid-state quantum technologies, hosting a variety of optically active defect centers with spin-selective transitions and emissions across visible to telecom wavelengths.[202]

The SiC exists in several polytypes (e.g., 4H, 6H, 3C), offering different crystallographic lattice configurations and defect environments.[203] This structural flexibility provides a rich platform for engineering diverse defect centers with tunable optical and spin properties. The SiC's wide bandgap (ranging from ~2.4 eV in 3C to ~3.2 eV in 4H) enables strong confinement of defect-related states within the bandgap, allowing stable emission and spin control at elevated temperatures and under electric fields.[204] Importantly, SiC is compatible with standard semiconductor doping and etching processes, enabling deterministic charge-state control, Stark tuning of optical transitions, and scalable nanophotonic integration.[205] The material's low nuclear spin environment (especially in isotopically purified forms) and low magnetic noise make it conducive to long spin coherence times, a prerequisite for quantum memory and quantum networking.[206]

**Key defect centers in SiC.** The SiC exhibits numerous defect centers, including but not limited to silicon vacancy ($V_{Si}$), divacancy ($V_{Si}V_C$), nitrogen vacancy ($V_{Si}N_C$), and transition metal-doped centers. The negatively charged silicon vacancy ($V_{Si}^-$) is an intrinsic point defect in SiC that possesses a quartet ground state ($S = 3/2$), distinguishing it from the more common spin-1 systems like the divacancy or NV centers.[207–210] This half-integer spin makes $V_{Si}^-$ robust against electric field perturbations and strain-induced decoherence, allowing stable quantum operations. In 4H-SiC, $V_{Si}^-$ can occupy hexagonal (h) and quasi-cubic (k) lattice sites, each yielding distinct ZPLs and spin transitions. The defect shows optical emission in the NIR (~860-950 nm) and supports ODMR-based spin initialization, manipulation, and readout at room temperature due to large ground-state zero-field splitting and strong magnetic field dependence.[211,212]

The $V_{Si}V_C$ divacancy, a complex of adjacent silicon and carbon vacancies, exhibits spin-1 ground states and coherent optical transitions near ~1100 nm.[213,214] It is structurally analogous to the NV center in diamond but offers longer spin coherence times under similar conditions. The divacancies in 4H-SiC exist in multiple configurations (hh, kk, hk, kh) depending on the crystal site and orientation.[206] The axial configurations (hh, kk) have higher symmetry, while the basal configurations (hk, kh) offer enhanced spin-orbit interactions.[214] Notably, divacancy centers at stacking fault interfaces exhibit improved spectral stability and narrow inhomogeneous broadening, attributed to environmental isolation.[215] These stacking-fault-hosted divacancies open new pathways for engineering site-specific defect placement using crystallographic imperfections, thereby relaxing the requirement for defect positioning in perfect lattices.[216]

The $V_{Si}N_C$ center offers a telecom-band analog of diamond's NV Centers. The $V_{Si}N_C$ complex in SiC consists of a silicon vacancy adjacent to a substitutional nitrogen atom replacing a carbon atom.[217] It shares similar spin and optical properties with the diamond NV center, but with a critical advantage of having ZPL lines in the NIR (1176–1243 nm), overlapping with the telecom O- and E-bands.[218,219] The defect exhibits short radiative lifetimes (2.1-2.8 ns) and ODMR-based spin control at room temperature.[218] While coherence times are typically shorter than divacancies, recent work demonstrates promising stability and tunability for long-distance quantum networks.

On the other hand, the extrinsic transition metal dopants like vanadium ($V^{4+}$) can also form deep-level defect states in SiC that act as single-photon emitters. The vanadium can substitute silicon sites during epitaxial growth or implantation, and in its charge state $V^{4+}$, it exhibits sharp emission in the 1300-1550 nm range with narrow linewidths and long-lived spin states.[203] These characteristics make vanadium centers promising for telecom-compatible spin-photon interfaces, although their spin control mechanisms remain less explored compared to intrinsic vacancies.

**Nanophotonic Integration and Device Applications.** The SiC's excellent etch ability and refractive index (~2.6) support the fabrication of high-quality nanophotonic structures such as waveguides, micro-ring resonators, and photonic crystal cavities. Integrating defect centers into these nanostructures enables Purcell-enhanced emission, improved collection efficiency, and spectral filtering.[220,221] Deterministic coupling between defect emitters and photonic modes has been demonstrated, with high beta-factors and emission directionality.[222] Moreover, SiC's compatibility with silicon-based electronics and photonics enables hybrid integration with CMOS technologies and cryogenic electronics for scalable control.[223]

SiC represents one of the most mature and industrially scalable hosts for solid-state spin-photonic qubits. Its diversity of defect centers, including intrinsic and extrinsic emitters spanning visible to telecom wavelengths, positions it as a flexible platform for quantum information processing and quantum networking.[224] Ongoing efforts in isotopic purification, deterministic defect placement, and hybrid integration are expected to further enhance its viability for on-chip quantum networks and spin-based quantum memories.[225,226]

### 3.4 Two-Dimensional Materials

The two-dimensional (2D) materials, atomically thin crystals with strong in-plane covalent bonding and weak out-of-plane van der Waals interactions, have introduced a new frontier in quantum photonics.[227] Their extreme spatial confinement, layer-dependent bandgaps, and large exciton binding energies enable rich light-matter interaction at room temperature, which is crucial for practical quantum photonic technologies.[228,229] Unlike traditional bulk semiconductors, 2D materials provide unique opportunities for controlling quantum emission through external perturbations, such as strain, electric fields, and dielectric environment.[230,231] These properties have enabled the development of highly localized, room-temperature single-photon sources in the visible and near-infrared (NIR) spectrum, with growing efforts toward telecom-band emission.[232,233]

The optical behavior of 2D semiconductors is dominated by tightly bound excitons, which are electron-hole pairs held together by Coulomb interaction. The exciton binding energies in transition metal dichalcogenides (TMDs), such as $WSe_2$, $WS_2$, and $MoS_2$, can exceed 300-700 meV, leading to stable excitonic emission even at room temperature.[234–236] These materials undergo a direct-to-indirect bandgap transition as they are thinned from bulk to monolayers. The monolayers exhibiting direct bandgaps in the visible-NIR range (e.g., 600-800 nm for $WSe_2$ and $MoS_2$).[237,238] The exciton recombination produces bright PL peaks in monolayers, typically dominated by neutral excitons, charged trions, and localized exciton states.[239] The spin-orbit coupling further splits the valence and conduction bands, giving rise to multiple excitonic transitions that can be spectrally resolved.[236] This spin-valley coupling forms the basis for potential spin-photonic qubit operations in 2D materials, although coherence control remains an active research area.

The localized excitons in 2D materials can exhibit photon antibunching, indicating single-photon emission. These emitters are often activated through defect formation, strain fields, or nanostructures, introducing localized potential wells that trap excitons.[240] The resulting emission is spatially confined, spectrally sharp (with linewidths <1 meV), and highly stable under ambient conditions. A hallmark study demonstrated single-photon emission from strain-induced QDs in monolayer $WSe_2$ using nanopillar substrates, with second-order correlation values $g^2(0) < 0.1$ and sub-nanosecond lifetimes.[241,242] The hexagonal boron nitride (hBN), a wide-bandgap insulator (~6 eV), also hosts ultra-bright room-temperature quantum emitters, though primarily in the visible regime (~550–750 nm).[243,244] These defect-based emitters exhibit high brightness, photostability, and nanosecond-scale lifetimes.[245] However, the large bandgap poses a challenge for telecom emission, motivating defect engineering strategies to create mid-gap states.[246]

**Strategies for Telecom-Band Emission.** Although most 2D quantum emitters operate in the visible spectral range, significant research focuses on bandgap engineering and adding defects in 2D materials to achieve single-photon emission in the telecom bands. The $MoTe_2$, a TMD with a narrow intrinsic bandgap (~1.0-1.1 eV), offers a particular example for bandgap engineering where the bandgap can be decreased by increasing layer thickness, enabling emission in the 1150-1300 nm range.[238,247] Furthermore, applying localized strain allows for additional redshift of the emission into the telecom C- and L-bands. While initial results have exhibited broad emission spectra, recent progress in site-controlled strain and heterostructure engineering is improving spectral purity and emission stability.[247] In parallel, defect engineering and artificial atom incorporation offer an alternative path. Introducing atomic-scale defects or embedding guest emitters, such as rare-earth ions, into 2D materials can create mid-gap electronic states that support optical transitions independent of the host's fundamental bandgap.[248] A notable example is erbium-doped $WSe_2$, which has demonstrated single-photon emission at 1530 nm with high spectral purity and desirable lifetime characteristics.[248] These techniques significantly broaden the accessible emission wavelengths of 2D materials. Additionally, interlayer excitons in van der Waals heterobilayers, such as $MoSe_2$-$WSe_2$ stacks, provide emission tunability via strain, and their long lifetimes make them attractive for electrically controllable NIR quantum light sources.[249]

**Nanophotonic Integration and Device Applications**. The 2D materials are suited for integration with nanophotonic structures due to their planar morphology, mechanical flexibility, and compatibility with various substrates.[250] Enhancing photon extraction, emission directionality, and spontaneous emission rates is crucial for overcoming their low quantum yield and inefficient out-of-plane emission. Coupling $WSe_2$ quantum emitters to silicon nitride photonic crystal cavities has yielded Purcell enhancement exceeding 10 times, significantly improving brightness and emission rates.[251] Similarly, plasmonic nanoantenna and dielectric metasurface have achieved subwavelength mode confinement, fast radiative decay with precise control over the emission spectra.[252] Mie resonators and plasmonic gratings integrated with hBN emitters have demonstrated orders-of-magnitude improvement in photon extraction efficiency, enhanced single-photon purity, and reduced background noise.[253,254] The deterministic placement techniques, such as nanopillar-induced strain localization and nanopore templating, enable emitter positioning with sub-100 nm precision, facilitating scalable integration into complex photonic circuits.[253,255] Recent developments in hybrid platforms further extend these capabilities by coupling 2D quantum emitters with fiber-based and CMOS-compatible photonic chips, paving the way toward wafer-scale, telecom-capable quantum light sources.[256]

Looking ahead, 2D materials offer a highly tunable and integrable platform for solid-state quantum photonics. While visible-range single-photon emission from $WSe_2$ and hBN are already well studied, the realization of spectrally narrow, stable, and telecom-band emitters remains a primary challenge. The emerging directions, such as strain engineering in narrow-bandgap TMDs like $MoTe_2$, rare-earth doping in wide-bandgap hosts like hBN, and the formation of interlayer excitons with electrical tunability, offer promising solutions. Combined with advances in nanophotonic integration, fiber coupling, and CMOS-compatible fabrication, 2D materials are poised to become key components of scalable room-temperature quantum networks.

## 4. Spin Photonic Qubits in Silicon

Silicon has recently emerged as a compelling host material for spin-photon qubits due to its unique ability to support optically active color centers, which exhibit long spin coherence times and emission in the low-loss telecom band (**Figure 2**).[54,257] Unlike other platforms such as diamond or SiC, which often require hybrid photonic integration, silicon color centers can be directly embedded within monolithic, CMOS-compatible photonic circuits.[258,259] The $^{28}Si$ isotope provides a magnetically isolated environment, eliminating most sources of spectral diffusion and spin dephasing, and enabling long coherence times approaching milliseconds. Several color centers have been identified in silicon, commonly known as the C center, $C_i$ center, G center, W center, and T center. [55,260–264] These defects, formed from common elements such as carbon, oxygen, and hydrogen, are accessible via standard fabrication techniques, such as ion implantation and thermal and laser annealing.[265–267] Importantly, their optical transitions span the telecom O-, C-, S, and L-bands, positioning silicon as an ideal platform for fiber-based quantum networks. This section reviews the properties and quantum functionality of these defects, beginning with the G center.

### 4.1 G Center (Carbon Pair Complex, ~1278 nm)

The G center is one of the earliest identified light-emitting defects in silicon and has become a key model system for quantum photonics in the material.[53,54] It consists of a triatomic carbon-carbon-silicon interstitial complex: two substitutional carbon atoms replace adjacent Si lattice sites, bound to a single silicon self-interstitial ($C_s$-$C_s$-$Si_i$).[265,268,269] This configuration has been confirmed through isotopic substitution experiments, where $^{13}C$ or different Si isotopes induce characteristic shifts in the optical spectrum. The G centers can be created via carbon implantation into silicon, followed by annealing at ~1000 °C, or by irradiation of carbon-containing silicon (e.g., Czochralski-grown) to generate mobile Si interstitials that combine with carbon pairs.[266,270] The defect forms readily in standard CMOS-compatible processing, and controllable positioning has been achieved using lithographic implantation masks.[264]

**Optical Properties and Spin Properties.** The G center's ZPL lies at ~0.969 eV (~1280 nm) in the telecom O-band, within the zero-dispersion window of standard optical fiber.[265] In natural silicon, inhomogeneous broadening of the ZPL (~0.1 meV) obscures fine structure.[271] In highly enriched $^{28}Si$, the linewidth narrows by more than two orders of magnitude, revealing a quartet of closely spaced lines corresponding to the four crystallographic orientations of the defect.[272] Ensemble linewidths as narrow as 0.4 μeV (~100 MHz) have been reported, approaching the transform limit given by ~6-8 ns radiative lifetime (natural linewidth ~20-30 MHz).[272] The single G centers typically exhibit linewidths of a few GHz in natural Si, but with photonic cavity stabilization, photons from a single emitter have shown HOM interference visibility.[273,274] Moreover, the DW factor is modest (~10-20% ZPL fraction), but reports of nanocavity coupling show >90% emission funneling into the ZPL.

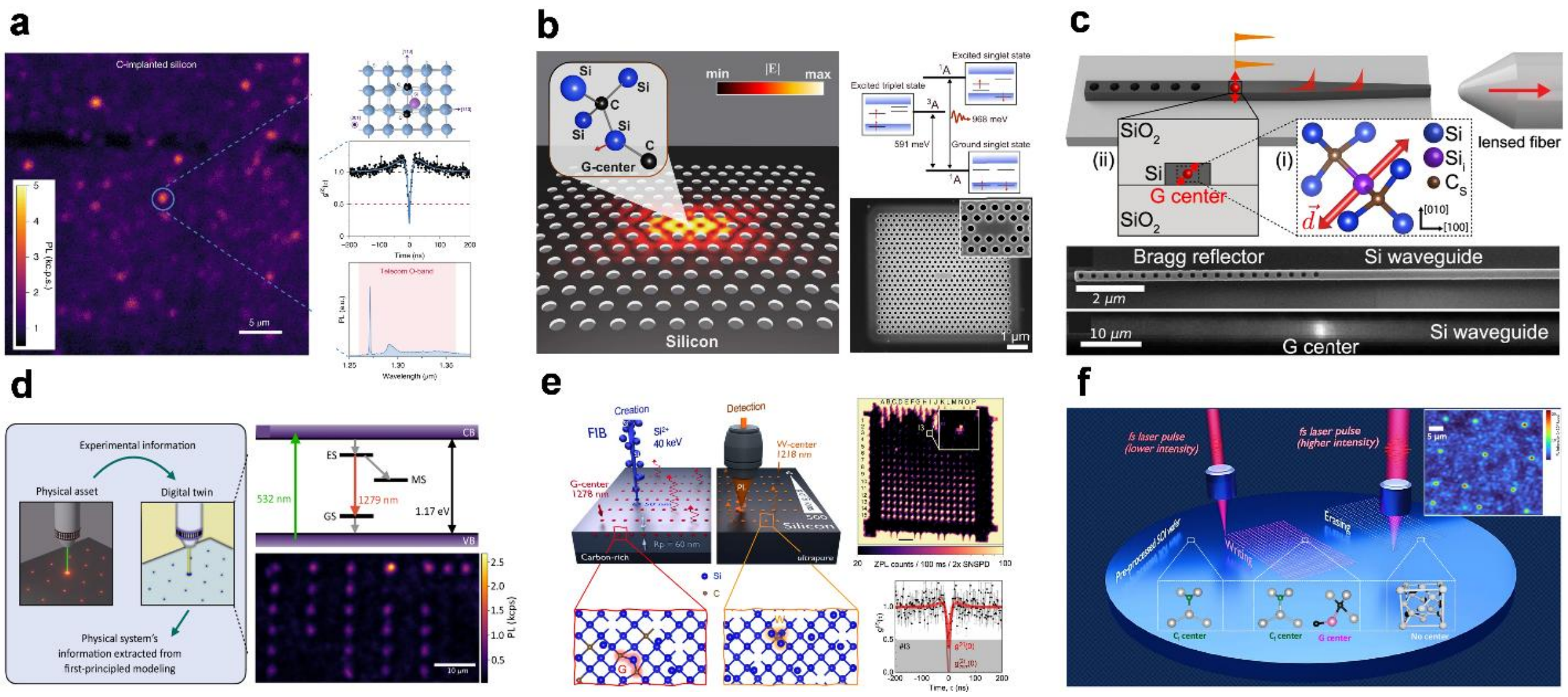

**Figure 7: Evolution of Color Centers in Si.** (a) First isolation and optical characterization of a single-color center in silicon. The photoluminescence (PL) spectrum from a carbon-implanted SOI wafer shows a sharp zero-phonon line (ZPL) at ~1278 nm corresponding to the G center ensemble. Inset: atomic structure of the G center, composed of two adjacent substitutional carbon atoms (black) bound to a silicon self-interstitial (purple). Antibunching at zero delay with $g^2(0) < 0.5$ confirms single-photon emission (Adapted from Ref. [54] by ©Springer Nature). (b) Demonstration of the first all-silicon quantum light source by embedding a single G center in a silicon nanocavity, achieving near-unity atom–cavity coupling, >30× Purcell enhancement, and >8× faster radiative rate. Inset: energy-level diagram of the G center, showing a singlet ground state and an optically excited singlet–triplet manifold (Ref. [278]; licensed under a Creative Commons Attribution (CC BY)). (c) Monolithic integration of G centers with silicon photonic waveguides. The SEM image shows a nanophotonic Bragg reflector used to collect indistinguishable single-photon emission from G centers into a lensed fiber (Adapted from Ref. [273]); licensed under a Creative Commons Attribution (CC BY)). (d) In-situ creation and activation of G centers in nanophotonic devices, demonstrating deterministic formation and reproducible optical characteristics. The inset shows the PL spectrum with identical fine-structure features across devices (Adapted from Ref. [313]; licensed under a Creative Commons Attribution (CC BY)). (e) Focused-ion-beam implantation of $Si^{2+}$ ions into carbon-rich silicon enables controllable fabrication of both G centers (1278 nm) and W centers (1218 nm). Inset: atomic model of the G center and the tri-silicon W complex. PL measurements confirm deterministic creation of these centers in predefined regions (Adapted from Ref. [264]; licensed under a Creative Commons Attribution (CC BY)). (f) Programmable creation and erasure of G and $C_i$ centers using femtosecond laser pulses in combination with hydrogen-based annealing and passivation. This laser-assisted defect engineering approach enables on-demand generation of telecom-band color centers, paving the way toward scalable integration of quantum emitters within CMOS-compatible silicon photonic platforms (Ref. [293]; licensed under a Creative Commons Attribution (CC BY)).

The G center was the first silicon color center isolated at the single-defect level (**Figure 7(a)**).[54,270] Under non-resonant excitation (e.g., 532 nm or above-band NIR), single centers emit stable emission in the telecom O-band, with $g^2(0) < 0.5$.[270] A very high emission rate has been reported.[257] The O-band offers low dispersion in fiber, making G-center photons attractive for quantum communication links. The high repetition rate allowed by the short lifetime (~10 ns) supports ~100MHz-clocked single-photon sources.

For decades, the G center was regarded as a spinless emitter as its ground state is a singlet ($S = 0$), with optical excitation forming a bound exciton and a short-lived metastable triplet state (~10-100 ns) lying below the bright singlet.[275,276] Early ensemble measurements in the 1980s detected ODMR linked to this triplet, but the effect was not observed at the single-defect level.[275] The view changed in 2025, when ODMR from a single G center integrated into a silicon bull's-eye cavity was demonstrated, along with coherent control of its electron spin.[277] The ODMR spectra revealed fine structure, potentially arising from different spin orientations due to defect center-of-mass motion.[272] Preliminary studies explored spin coherence times, and the results suggest that nearby $^{13}C$ or $^{29}Si$ nuclear spins could be hyperfine-coupled to the triplet electron, enabling nuclear-spin quantum memories. The combination of an electron spin-free ground state with the ultra-low magnetic noise of $^{28}Si$ offers the prospect of exceptionally long nuclear coherence times, similar to donor nuclear spins in silicon.[269] The demonstration of single-defect ODMR and coherent spin control in 2025 redefines the role of the G center from a "spinless" emitter to a potential spin-photon interface.

**Integration with Silicon Photonics.** The G center is fully CMOS-compatible as it comprises only silicon and carbon. It has been integrated into silicon photonic crystal cavities, ring resonators, and waveguides, with deterministic placement enabling optimal dipole-cavity alignment (**Figure 7(b)-(e)**). [274,277–280] The Purcell factors >30 have also been achieved, boosting ZPL emission and radiative rate by nearly an order of magnitude.[278] In addition, the G centers have been coupled directly incorporated into silicon LEDs emitting at 1278 nm, raising the prospect of electrically driven single-photon sources **(Figure 8(e)-(f)).**[281,282] Its telecom O-band photons can be highly indistinguishable, and coupling to nearby nuclear spins could provide long-lived quantum memories. Combined with its exceptional integration compatibility and demonstrated cavity/waveguide coupling, the G center is now positioned as a versatile building block for silicon-based quantum networks, serving as a bright single-photon source and, potentially, as part of a spin-photon-nuclear memory architecture.

### 4.2 T Center (Carbon-Hydrogen Complex, ~1326 nm)

The T center is a spin–photon color center in silicon formed from a carbon–hydrogen complex, typically described as $(C\text{–}C\text{–}H)_{Si}$, where two carbon atoms occupy a substitutional site, one terminated by hydrogen.[55,283] The prevailing formation model involves an interstitial carbon first binding to hydrogen (forming C–H), then migrating to pair with a substitutional carbon, producing a monoclinic-I ($C_{1h}$) symmetry defect.[284] The T center forms in carbon- and hydrogen-containing silicon following irradiation and annealing at 350–600 °C, with hydrogen introduced via forming-gas anneals, plasma treatment, or proton implantation.[285,286] Excess hydrogen can passivate the defect, emphasizing the need for precise process control.[287] The center has been observed in float-zone, Czochralski, and SOI silicon, and can be created post-fabrication in photonic devices through targeted implantation and annealing.[112,288,289] Its identity is confirmed by isotope-dependent optical shifts under $^{13}C$ and deuterium substitution. The defect's simple composition and compatibility with CMOS processing make it readily integrable within foundry-scale silicon photonic platforms (**Figure 8(a–d)**).

**Optical and Spin Properties.** The T center exhibits a zero-phonon-line (ZPL) doublet near 0.935 eV (~1326 nm, telecom O-band), split by ~1.8 meV into two excitonic transitions ($TX_0$ and $TX_1$) due to spin–orbit coupling and local strain.[263,290] At cryogenic temperatures, emission is dominated by the lower-energy $TX_0$ line.[55] In natural silicon, inhomogeneous broadening from $^{29}Si$ nuclear spins and strain leads to linewidths of 0.1–0.3 meV and a Debye–Waller (DW) factor of ~3%.[55] In isotopically purified $^{28}Si$, the absence of spin noise and disorder narrows ensemble linewidths to ~0.25 μeV (~60 MHz)—approaching the lifetime limit set by the ~0.94 ns radiative lifetime—and the DW factor rises to ~23%.[291] The phonon sideband extends toward 1.5 μm, but in $^{28}Si$ a substantial portion of emission remains concentrated within the ZPL doublet. Single-defect spectra exhibit excellent spectral stability and minimal diffusion below 4 K, especially in low-strain or uncladded photonic environments, underscoring the center's suitability for integration into silicon nanophotonics.

The neutral T center hosts an unpaired electron spin (S = ½) hyperfine-coupled to the bonded $^{1}H$ nuclear spin (I = ½), forming an intrinsic two-qubit register (**Figure 8(a)**).[93] A nearby $^{29}Si$ nuclear spin (I = ½) can act as a third qubit, enabling multi-qubit operation within a single atomic defect.[93] The electron spin exhibits an almost isotropic g-factor of ~2.0055, and hyperfine coupling with $^{1}H$ produces a resolvable splitting of ~117 MHz for specific crystallographic orientations.[55,93] Optical transitions are primarily spin-conserving, permitting polarization-selective spin initialization and high-fidelity single-shot readout via resonant excitation—analogous to protocols in NV centers in diamond. At 1.2 K, measured coherence times reach $T_2 \approx 0.41$ ms for the electron spin, $T_{2,H} \approx 112$ ms for the $^{1}H$ nuclear spin, and $T_{2,Si} \approx 67$ ms for a coupled $^{29}Si$ nuclear spin.[93] These coherence values rival or exceed many solid-state spin qubits and can be extended further through $^{28}Si$ isotopic enrichment and dynamical decoupling. The ability to stably host multiple long-lived qubits within a single defect—together with direct optical transitions in the telecom band—makes the T center a uniquely powerful candidate for on-chip quantum memories and entanglement-based communication nodes.

**Emission, Photonic Integration, and Quantum Network Potential**. Single T centers exhibit strongly antibunched emission ($g^2(0) \rightarrow 0$) in the telecom O-band, fully compatible with low-loss fiber and integrated waveguides.[55,93] Integration with high-Q photonic crystal cavities yields >10× enhancement of ZPL emission and >10% on-chip collection efficiency.[93] Electrically driven electroluminescence from T centers has been realized in p–i–n diodes, enabling on-demand single-photon generation without external optical excitation (**Figure 8(d)**).[292] Post-fabrication creation in nanobeams, rings, and waveguides is routine, while cavity coupling provides Purcell enhancement to improve brightness, coherence, and indistinguishability.[291]

The combination of telecom-band emission, nuclear spin memory, and silicon photonic compatibility uniquely positions the T center for scalable quantum networking. The long-lived nuclear spins act as quantum memories, while the electron spin mediates spin–photon entanglement for inter-node communication. In prospective repeater nodes, T centers embedded in photonic circuits could distribute entanglement through optical fibers while storing quantum information locally for milliseconds to seconds. Unlike NV centers, no frequency conversion is required, and unlike SiC divacancies, the silicon host can be isotopically purified to suppress spin noise. Although cryogenic operation ($< 4$ K) is required, the compatibility with cryo-CMOS technology makes such integration realistic for large-scale systems.

The T center is the first silicon defect to demonstrate full, on-chip multi-qubit control, bridging electron and nuclear spin registers with photonic interfaces. In a landmark experiment, Song et al. demonstrated a three-qubit register within a silicon photonic device comprising the electron spin, the intrinsic $^{1}$H nuclear spin, and a nearby $^{29}$Si nuclear spin (**Figure 8(b)**).[93] Resonant optical pulses at the ZPL were used for spin-selective initialization and single-shot readout, while microwave and radio-frequency fields achieved coherent spin rotations. This enabled high-fidelity single-qubit gates on all three spins and a two-qubit nuclear–nuclear entangling gate, mediated by the electron spin, with ~77% fidelity. The resulting Bell states between nuclear spins exhibited coherence lifetimes of several milliseconds, reflecting the remarkable stability of nuclear spin memories in silicon. The electron spin simultaneously serves as the optical interface, enabling quantum state mapping between nuclear spins and telecom photons—an essential capability for quantum repeaters and distributed entanglement protocols.

Overall, the T center represents the current state of the art in silicon spin–photon interfaces—a solid-state system uniting long spin coherence, intrinsic multi-qubit registers, telecom-band optical transitions, and CMOS process compatibility. Recent demonstrations of on-chip three-qubit entanglement highlight its readiness for integration into scalable, fiber-connected quantum computing and networking architectures.[93] With deterministic placement, isotopically pure hosts, and integrated control, the T center is poised to become the foundational platform for silicon-based quantum repeaters and distributed quantum processors.

### 4.3 $C_i$ Center (Carbon Interstitial, ~1452 nm)

The $C_i$ center is a point defect in silicon consisting of a single carbon atom at an interstitial site. It is structurally modeled as a split-interstitial $(Si\text{-}C)_{Si}$ pair where a carbon atom shares a lattice position with a displaced silicon atom.[56,293] Unlike the G center (two carbons) or the C center (C-O pair) (discussed later), the $C_i$ contains only one carbon atom with no additional light atoms. It can be reliably formed by carbon ion implantation followed by thermal annealing. However, hydrogen co-doping is critical where annealing in forming gas ($H_2/N_2$) strongly promotes $C_i$ formation while suppressing competing complexes, likely by passivating dangling bonds and stabilizing the split-interstitial geometry.[293] For instance, implanting $^{12}C$ or $^{13}C$ into SOI wafers and annealing at ~800 °C in forming gas yields high densities of $C_i$ centers with emission at ~1452 nm, while minimizing G center formation (**Figure 7(f)**). Femtosecond-laser-based "write/erase" methods have enabled programmable activation and quenching of $C_i$ centers by manipulating hydrogen motion, allowing deterministic placement of single emitters directly in prefabricated photonic circuits.

**Optical and Spin Properties.** The $C_i$ center exhibits a zero-phonon line (ZPL) near 0.854 eV (~1450–1452 nm), positioned within the extended telecom band. Although historically less studied, it has recently attracted significant attention owing to its exceptionally narrow linewidth and full CMOS compatibility. In hydrogen-annealed SOI devices, individual $C_i$ centers display ZPL linewidths as narrow as 0.03 nm (~4.2 GHz), limited by spectrometer resolution.[293] This remarkable spectral stability is attributed to hydrogen passivation, which mitigates strain and charge noise at the $Si/SiO_2$ interface. Forming-gas annealing allows hydrogen to diffuse into the silicon lattice, neutralizing dangling bonds and compensating interfacial stress—thereby suppressing local electric-field fluctuations and lattice distortions that typically cause spectral diffusion. As a result, the $C_i$ center shows significantly reduced inhomogeneous broadening compared to other silicon color centers. Time-resolved photoluminescence reveals fast radiative lifetimesnof 3–8 ns, corresponding to transform-limited linewidths in the tens of MHz and enabling highly indistinguishable photon emission in isotopically enriched $^{28}Si$. While the Debye–Waller factor remains moderate, integration with high-Q nanocavities can funnel emission into the ZPL, enhancing brightness and photon collection efficiency for on-chip quantum photonics. The electron paramagnetic resonance studies have shown that the $C_i$ defect possesses a paramagnetic ground state with S=1/2 in certain charge configurations.[294] Two optically addressable spin-1/2 charge states appear to exist,[56] making the $C_i$ a promising spin-photon qubit candidate. Although single-defect ODMR has not yet been demonstrated, the analogy to the T center suggests that optical spin polarization and readout should be achievable with resonant excitation. The hyperfine interactions with the $^{13}C$ nucleus or nearby $^{29}Si$ can provide a route to nuclear spin quantum memories. Moreover, the optically active state is likely neutral or negative, and charge stability can be engineered via local doping or biasing. While no $T_2$ measurements exist, the absence of a strong Jahn-Teller distortion and the precedent of long coherence in other carbon-based silicon defects suggest that millisecond-scale $T_2$ times could be achievable in isotopically purified silicon.[295]

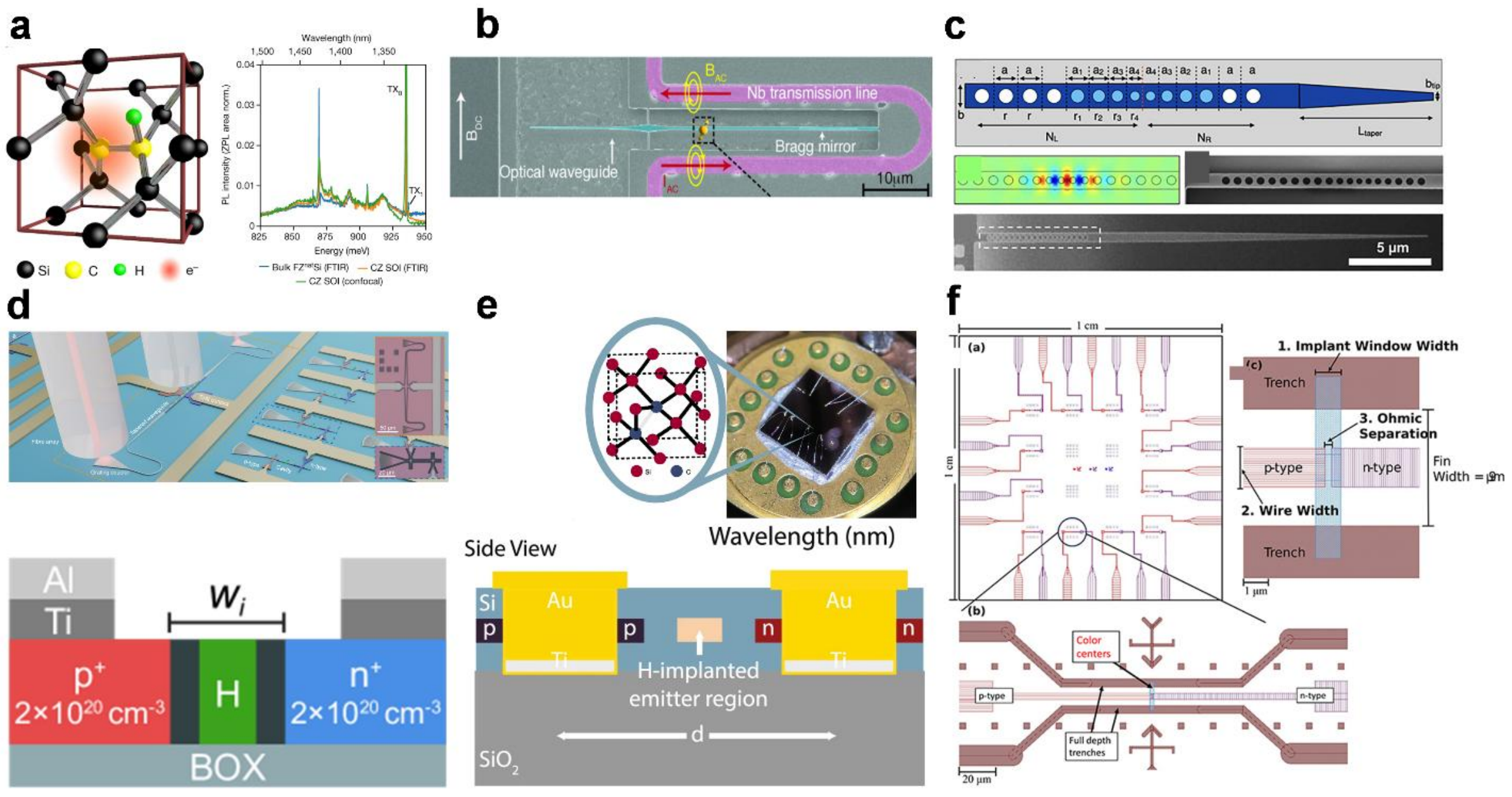

**Figure 8: Optically and electrically triggered spin-photon interface in silicon:** (a) Atomic structure of spin-dependent color center in Si, T center, demonstrate optical transitions at telecom band wavelengths (1326 nm). Individually addressable T center integrated with silicon photonic structures consists of an unpaired electron spin and a hyperfine-coupled hydrogen nuclear spin (Adapted from ref. [263] by ©Springer Nature). (b) Initialization and state read-out of long-lived three-qubit register with an optical interface based on electron spin of T center coupled to a hydrogen and silicon nuclear spin. These long-lived nuclear spins (~112 ms for hydrogen nuclear spin and ~67 ms for silicon nuclear spin) of T center in a silicon photonic waveguide can serve as quantum memory nodes for scalable quantum network along with three qubit registers (Adapted from ref. [93]; licensed under a Creative Commons Attribution (CC BY)). (c) Integrating optically active spin qubit of T center with nanophotonic cavities enhance emission characteristics especially brightness and collection efficiency from quantum emitter to fiber paves the way to efficient spin-photon interface in silicon (Adapted from ref.[291] by © ACS). (d) Electrically triggered single photon emission from T center in Si is integrated with an optoelectronic device that combines nanophotonic waveguides and p-i-n diodes. Optoelectronic device configuration to coherently control electron spin of T center makes it a versatile defect center for scalable quantum technology (Adapted from ref. [292] © Springer Nature). (e) Electrical manipulation in ensemble G centers with lateral p+-p-n+ diodes fabricated in commercial SOI wafer provides broad applicability for color centers in Si (Adapted from ref. [281]; licensed under a Creative Commons Attribution (CC BY)). (f) Fabricating light-emitting-diodes with G center and W center in silicon demonstrates versatility of color centers in Si for scalable quantum technologies (Adapted from ref. [282] by ©AIP Publishing).

**Emission and Integration.** Recent experiments have isolated single $C_i$ centers with controlled positioning using focused femtosecond pulses, achieving ~50% yield and producing clean single-photon emission ($g^2(0) \approx 0$) at ~1452 nm.[293] The ability to post-fabrication "write" or "erase" emitters enables defect placement aligned to photonic modes in a photonic nanostructure, a unique advantage over most other solid-state platforms. The $C_i$ centers can be implanted directly into the device layer of SOI wafers, coupling efficiently to silicon waveguides. They show distinct strain responses compared to G centers, allowing frequency tuning via local stress or electric fields. The ~1452 nm photons propagate in low-loss Si waveguides and can be routed to on-chip filters, ring resonators, or superconducting detectors. Although spin-based quantum operations have not yet been realized in isolated centers, the combination of a telecom ZPL, narrow linewidth, and EPR-verified spin-1/2 states makes the center a strong candidate for future spin-photon interfaces, two-photon interference experiments, and integration into Purcell-enhanced nanocavities for bright, indistinguishable single-photon generation.

### 4.4 C Center (Carbon-Oxygen Complex, ~1571 nm)

The C center in silicon is a well-characterized isoelectronic bound exciton complex formed by an interstitial carbon paired with an interstitial oxygen.[34,296,297] This defect binds an exciton that produces a sharp ZPL at 1571 nm, well within the telecom L-band.[34,297–299] It can be created in Czochralski-grown silicon via electron or proton irradiation followed by annealing, which mobilizes interstitial carbon and oxygen atoms to form the complex.[57,300]

**Optical and Spin properties:** The optical transition corresponds to radiative recombination of the bound exciton, giving rise to a prominent and stable emission line known as the C-line. In natural silicon, the ZPL is broadened by isotopic disorder, but in isotopically purified $^{28}Si$, the inhomogeneous linewidth narrows dramatically to the μeV scale, approaching the transform-limited value.[299] These ultra-narrow linewidths, well-defined emission energy, and minimal spectral diffusion make the C center a promising candidate for generating indistinguishable photons suitable for quantum interference and entanglement distribution.

The C center possesses a singlet ($S = 0$) ground state and triplet ($S = 1$) excited state.[34,57] This excited-state triplet enables optical spin access, despite the non-paramagnetic ground state. Recently, ODMR was demonstrated in ensembles of C centers, confirming the ability to manipulate and read out spin populations using microwave at telecom wavelengths. The ODMR spectra revealed zero-field splitting between the $m_s = \pm 1$ levels of the triplet state (~44 MHz), in agreement with theoretical predictions.[34] The triplet state exhibits nonradiative lifetimes exceeding 10 ms, allowing long-lived spin population storage.[57] The results mark the first demonstration of spin-dependent optical emission near telecom C-band wavelength in silicon, opening the door to quantum memory and entanglement protocols within lowest loss existing fiber-optic infrastructure.

**Emission and Integration.** While single C centers have not yet been fully isolated, the narrow homogeneous linewidths in $^{28}Si$ suggest the possibility of high photon indistinguishability and negligible spectral diffusion. Being a bound exciton emitter, the DW factor is moderate, with a portion of emission occurring in phonon sidebands. However, integration into high-Q photonic cavities can enhance emission into the ZPL via the Purcell effect. The triplet sublevels may further enable spin-photon entanglement schemes, where photon polarization or frequency correlates with spin state. Although such protocols are still in early stages for C centers, the long-lived triplet state, potential for nuclear spin coupling (e.g., to $^{13}C$ nuclei), and transform-limited linewidths position them as viable components in scalable quantum networks.

From a fabrication standpoint, the C center is highly compatible with standard silicon processing. Its constituents, carbon and oxygen, are common impurities, and formation via ion implantation and annealing enables deterministic placement post-fabrication.[57] The telecom emission wavelength allows direct coupling to fiber networks with minimal insertion loss. Together with the demonstration of ODMR and telecom-band emission, these characteristics make the C center one of the most promising spin-photonic qubit candidates in silicon.

### 4.5 Erbium in Silicon (~1540 nm)

Erbium ($Er^{3+}$) is a trivalent rare-earth ion whose intra-4f shell transition $^4I_{13/2} \rightarrow {}^4I_{15/2}$ produces a sharp emission line at ~1540 nm, matching the optical fiber's low-loss C-band.[301] In crystalline silicon, Er substitutes for a silicon atom, often accompanied by a charge-compensating defect such as an interstitial oxygen (forming Er-O complexes) to enhance optical activation.[302] Because filled 5s and 5p orbitals strongly shield the 4f electrons, their optical transitions are largely insensitive to the host lattice, yielding narrow inhomogeneous linewidths even in an amorphous or strained environment. The Er can be incorporated into silicon through ion implantation, molecular beam epitaxy, or Czochralski growth with in-situ doping.[303] A post-implantation annealing at ~800-900 °C repairs implantation damage and promotes Er-O pairing, significantly increasing emission efficiency. The incorporation process is CMOS-compatible, enabling Er integration into standard SOI photonics platforms.[304]

**Optical and Spin Properties.** Erbium's 4f-4f transitions are parity-forbidden electric-dipole transitions, weakly allowed via mixing with higher-lying states, leading to long excited-state lifetimes (ms range) and extremely narrow homogeneous linewidths at cryogenic temperatures (down to the kHz range).[303] The ZPL transition at ~1.54 μm is accompanied by weak phonon sidebands, with a very high DW factor (>90%), allowing nearly all emission to be concentrated in the ZPL.[305] The long lifetime (~1-10 ms in crystalline silicon at cryogenic temperatures) allows for high-fidelity quantum state storage in the optical domain. Still, it limits the photon emission rate unless enhanced via the Purcell effect using nanophotonic cavities. The cavity coupling has achieved enhancement factors exceeding 50x, enabling single-ion emission rates approaching MHz levels.[13]

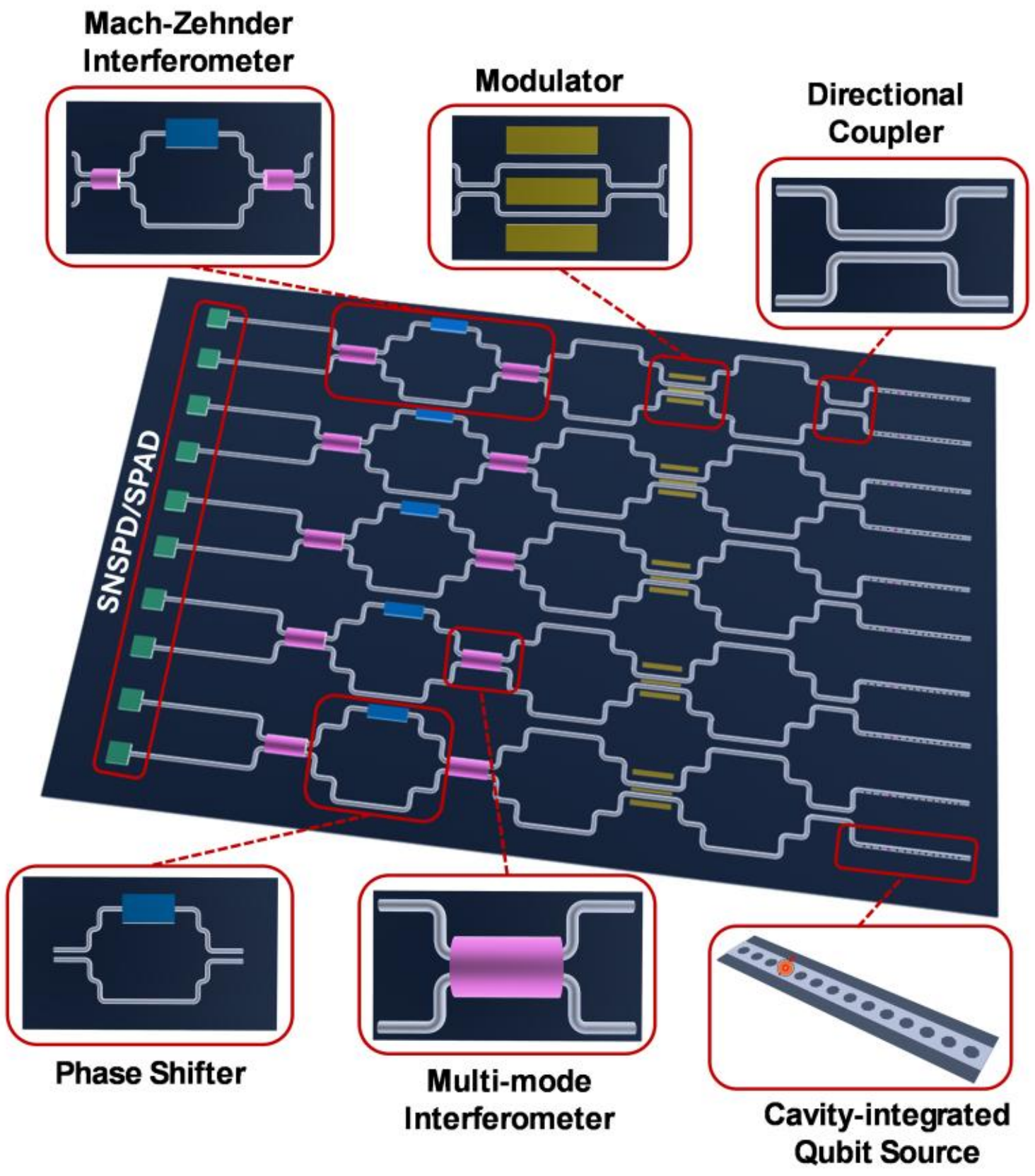

**Figure 9: Vision for a Monolithic Silicon Quantum Photonic Integrated Circuit.** Schematic representation of an all-silicon quantum photonic integrated circuit (QPIC) that integrates optically addressable or electrically injected spin–photon qubits with CMOS-compatible silicon photonics. Key photonic components, including Mach–Zehnder interferometers, phase shifters, directional couplers, and multi-mode interferometers, enable on-chip quantum operations such as Hong–Ou–Mandel (HOM) interference, Hanbury Brown and Twiss (HBT) correlation measurements, and Bell-state projections. Spin-photonic qubits, such as silicon color centers (e.g., T, C, and Ci centers), are embedded within nanophotonic cavities or waveguides (zoomed-in panel) to enhance emission via the Purcell effect and improve photon extraction into guided modes. On-chip single-photon detection is achieved through integrated superconducting nanowire single-photon detectors (SNSPDs) or Ge/Si single-photon avalanche diodes (SPADs), while future sub-bandgap Si-SPADs could enable room-temperature, fully monolithic photon detection. Programmable phase and amplitude control provided by modulators allow dynamic reconfiguration of quantum circuits for scalable operations. Together, such integrated functionality paves the way for on-chip spin–photon entanglement, quantum logic operations, and distributed quantum network nodes within a unified silicon platform.

$Er^{3+}$ in silicon has an effective electron spin arising from its 4f manifold, with the $^{4}I_{15/2}$ ground state split into Stark sublevels by the local crystal field.[306] Certain isotopes, notably $^{167}Er$ (natural abundance ~23%), possess a nuclear spin I=7/2, enabling hyperfine-resolved optical and spin transitions.[307] This combination allows for long-lived spin and hyperfine states, which can be optically addressed through the telecom transition. The electron g-factor is typically in the range 2-8. At low magnetic fields, spin-lattice relaxation times $T_1$ can reach seconds, and coherence times $T_2$ for the electron spin have been measured up to milliseconds, with nuclear spin $T_2$ extending into the second range under dynamical decoupling.[305,307] Optical transitions can be spin-selective, enabling Λ-schemes for optical spin initialization and readout.

The $Er^{3+}$ is one of the few solid-state emitters directly combining long-lived optical and spin transitions in the telecom band.[306] This has enabled demonstrations of spin-photon entanglement, optical-spin quantum memories, and photon-spin-photon teleportation in other host materials, and similar protocols are being extended to silicon.[308,309] In isotopically enriched $^{28}Si$, single-ion optical linewidths have been observed at the lifetime limit, and photon-echo experiments have yielded optical coherence times $T_2$ exceeding hundreds of microseconds.[307] The hyperfine states of $^{167}Er$ serve as long-lived quantum memories, with limited coherence primarily due to residual magnetic noise.[307]

**Emission and Integration.** $Er^{3+}$ emission wavelength perfectly matches the existing fiber infrastructure, eliminating the need for frequency conversion. Er ions can be integrated into SOI photonic crystal cavities, micro-ring resonators, and waveguides with high yield, enabling efficient fiber-to-chip coupling. Recent advances have demonstrated single-ion detection in silicon nanophotonic cavities with Purcell factors >100, enabling high signal-to-noise single-photon generation at 1.54 μm.[310] The electrical control of $Er^{3+}$ optical transitions via the Stark effect has been achieved, allowing frequency tuning and spectral alignment of multiple ions for interference-based protocols.[311] The combination of telecom-band emission, long-lived spin and optical coherence, and CMOS process compatibility makes $Er^{3+}$ in silicon a compelling platform for quantum repeaters, memory-assisted entanglement distribution, and hybrid integration with superconducting or spin-based quantum processors.[52]

**Vision for Monolithic Silicon Quantum Photonic Circuits for Scalable and Distributed Quantum Networks.** Among silicon-based telecom-wavelength emitters, each system offers a distinct balance between optical performance, spin properties, and integration maturity. The T center stands out as the most advanced optically addressable spin-photon interface in silicon, enabling multi-qubit registers and spin-photon entanglement, but suffers from a relatively low DW factor (~3-5 %). The G center offers bright, transform-limited O-band photons and now supports ODMR at the single-defect level. Yet, its non-paramagnetic ground state necessitates exploiting a metastable triplet for spin control. The $C_i$ center combines simplicity of composition with extended-telecom emission (~1452 nm), less inhomogeneous broadening, and narrow linewidths in nanophotonic environments, while single spin readout remains unexplored. Its $S = 1/2$ ground state and compatibility with laser writing make it a promising platform for programmable on-chip networks. The C center delivers ultra-narrow L-band photons and ensemble ODMR but has not yet been demonstrated at the single-defect level. In contrast, $Er^{3+}$ in silicon offers exceptional coherence and high DW factors, making it an ideal telecom-band quantum memory. Together, these platforms define a rich design space for scalable, CMOS-compatible quantum photonic architectures, where the optimal choice depends on whether the priority is high-brightness indistinguishable photons, long-lived quantum memories, or integrated multi-qubit control.

The convergence of silicon spin-photon qubits, integrated photonics, and on-chip single-photon detectors offers a transformative path toward scalable quantum networks and photonic quantum computing. As envisioned in **Figure 9**, monolithic QPICs can co-integrate deterministic spin-photon sources, such as telecom-band silicon color centers, with low-loss silicon waveguides, reconfigurable interferometric circuits, and high-performance detectors. Coherent spin-photon interfaces enable on-chip generation of indistinguishable single photons entangled with long-lived electron or nuclear spins. At the same time, integrated Mach-Zehnder interferometers, phase shifters, and modulators provide fast and programmable photonic processing. Detection can be realized via superconducting nanowire single-photon detectors (SNSPDs) or Ge/Si SPADs, and future advances in sub-bandgap absorption-based silicon SPADs could yield fully CMOS-compatible, cryo-to-room-temperature telecom-band detection directly on chip. Such architecture allows compact implementation of quantum communication protocols, Bell-state measurements, HOM interference, entanglement swapping, and seamless interfacing with fiber networks for distributed quantum systems. By leveraging the silicon photonics platform's scalability, stability, and fabrication maturity, these QPICs can serve as universal quantum photonic interconnect modules and quantum network nodes, enabling both large-scale photonic quantum computing and fault-tolerant distributed quantum networks.

## 5. Conclusion and Outlook

Solid-state quantum emitters have advanced rapidly over the past two decades, offering diverse routes toward scalable quantum technologies. Diamond color centers, epitaxial quantum dots (QDs), defects in silicon carbide (SiC), and emerging two-dimensional (2D) material emitters have each demonstrated essential quantum functionalities—long spin coherence, high-purity single-photon emission, and integration with photonic nanostructures. These systems have enabled landmark demonstrations in quantum communication, sensing, and computation. However, each platform faces trade-offs between emission wavelength, scalability, and fabrication complexity.

Silicon defect centers and dopants now occupy a unique and promising niche in this landscape. Their intrinsic telecom-band optical transitions enable direct compatibility with existing fiber infrastructure, while the silicon host benefits from decades of industrial maturity in CMOS processing. Recent breakthroughs—including coherent spin control of single G centers, multi-qubit spin registers in T centers, transform-limited linewidths in Ci centers, and optically detected magnetic resonance in C centers—highlight the rich quantum functionalities achievable in silicon. Integration with advanced silicon photonics further enables low-loss routing, high-Q cavity enhancement, and scalable on-chip quantum operations. Complementary progress in on-chip single-photon detectors—superconducting nanowire (SNSPDs), Ge/Si SPADs, and emerging sub-bandgap Si SPADs—completes the monolithic architecture required for all-silicon quantum photonic platforms. Looking forward, integrating spin–photon qubits with reconfigurable photonic circuits, quantum frequency conversion, and multiplexed single-photon detection will be central to realizing fully functional quantum processors and network nodes.

The long-term vision is a monolithic silicon quantum photonic platform where arrays of optically addressable spin qubits are seamlessly integrated with on-chip quantum photonic logic and high-efficiency detectors. Such an architecture would enable scalable photonic quantum computing, metropolitan-to-global quantum networking, and distributed quantum sensing—all within the material backbone of modern microelectronics. This convergence of defect engineering, nanophotonics, and CMOS fabrication positions silicon as a cornerstone of the emerging quantum internet.

From a broader technology perspective, each material system plays a distinct and complementary role: diamond and SiC color centers lead to long-distance quantum repeaters owing to their exceptional spin coherence; QDs provide ultrafast deterministic photon sources for photonic processors; and 2D materials enable compact, tunable quantum light emitters. Silicon, however, stands out as the natural hub for large-scale integration, linking telecom-compatible spin–photon interfaces with dense photonic circuits and integrated detectors—forming the backbone of future global quantum networks. While all-silicon quantum photonic systems can serve as universal interconnect modules in heterogeneous quantum architectures, silicon spin–photon qubits also offer a homogeneous and intrinsically scalable route to unified quantum processing, memory, and networking on a single chip.